\documentclass[11pt]{article}

\usepackage[table,xcdraw]{xcolor}
\usepackage[utf8]{inputenc}
\usepackage[margin=3cm]{geometry}
\usepackage{graphicx}
\usepackage{setspace}
\usepackage{amsmath}
\usepackage{amssymb}
\usepackage{url}
\usepackage{longtable}
\usepackage{pdflscape}
\usepackage{hyperref}
\usepackage{tikz}
\usepackage{tabularx}
\usepackage{booktabs}
\usepackage{makecell}
\usepackage{titlesec}
\usepackage[defaultlines=2,all]{nowidow}
\usepackage[margin=2cm,font=footnotesize]{caption}
\usepackage{subcaption}
\usepackage{nameref}
\usepackage{dcolumn}
\usepackage{rotating}
\usepackage[export]{adjustbox}
\usepackage{cleveref}
\usepackage{csquotes} 

\usepackage{float}
\usepackage{multirow}
\usepackage{amsthm}
\usepackage{mathrsfs}
\usepackage[title]{appendix}
\usepackage{textcomp}
\usepackage{manyfoot}
\usepackage{algorithm}
\usepackage{algorithmicx}
\usepackage{algpseudocode}
\usepackage{listings}

\usetikzlibrary{positioning, arrows.meta, calc}

\titlespacing{\paragraph}{0pt}{0pt}{1ex}

\renewcommand{\epsilon}{\varepsilon}

\usepackage[backend=biber,style=authoryear,
citestyle=authoryear,isbn=false,url=false,eprint=false, dashed=false,
giveninits=true,maxcitenames=1,uniquename=init,natbib=true]{biblatex}
\renewcommand{\cite}{\parencite}

\AtEveryBibitem{
  \clearfield{note}
  \clearfield{url}
  \clearfield{issn}
  \clearfield{urldate}
  \clearfield{month}
}

\definecolor{bleucite}{RGB}{34,111,212}
\hypersetup{hidelinks}
\AtEveryCite{\color{bleucite}}

\title{Link Fraction Mixed Membership Reveals Community Diversity in Aggregated Social Networks}

\author{Gamal Adel$^{1,*}$, Eszter Bokányi$^1$, Eelke M. Heemskerk$^2$, Frank W. Takes$^1$}

\date{{\footnotesize%
     $^1$Leiden University, The Netherlands $^2$University of Amsterdam, The Netherlands\\[1ex]
     $^*$\texttt{elgamali@leidenuniv.nl} (Corresponding author)\\[2ex]
\today}}

\begin{document}

\maketitle

\onehalfspacing

\begin{abstract}

Community detection is a critical tool for understanding the mesoscopic structure of large-scale networks. However, when applied to aggregated or coarse-grained social networks, disjoint community partitions cannot capture the diverse composition of community memberships within aggregated nodes. While existing mixed membership methods address this issue, they often result in community assignments that are inconsistent across varying scales of aggregation.
This paper presents the Link Fraction Mixed Membership (LFMM) method, which computes the mixed memberships of nodes in aggregated networks. Unlike existing mixed membership methods, LFMM is consistent under aggregation. Specifically, we show that it conserves community membership sums at different scales.
The method is utilized to study a population-scale social network of the Netherlands, aggregated at different resolutions. Experiments reveal variation in community membership across different geographical regions and its evolution over the last decade. In particular, we show how our method identifies large urban hubs that act as the melting pots of diverse, spatially remote communities.
\end{abstract}

\section{Introduction}

One of the most characteristic properties of large-scale social networks is their community structure, as it can reveal social tendencies and association patterns within a population at the mesoscale level~\cite{backstrom2006group}. Communities, groups of individuals that are more connected to each other than with other groups, can provide insight into social network mechanisms such as homophily~\cite{Menyhert2024ConnectivityAC}, segregation~\cite{kazmina2024socio}, and information spread~\cite{mantzaris2014uncovering}. However, evaluating the community structure of individual-level social networks is often untenable, as such data may be inaccessible due to privacy concerns or being too computationally expensive to process~\cite{peng2018social,de2024effect}.

As a common alternative, aggregated (coarse-grained) networks are typically constructed by partitioning nodes into disjoint sets (e.g., through clustering or grouping by geographical and affiliation attributes) and then summing up edge counts or weights between nodes in these sets~\cite{Kim2004GeographicalCG}. 
While aggregation destroys much of the individual-level information of the network, it has been shown that it can give mesoscopic insights about the underlying network~\cite{butts2009revisiting}. However, there is currently a lack of tools specifically designed for properly analyzing aggregated networks, resulting in many existing works in the literature simply treating them as weighted networks~\cite{Decuyper2018MeasuringTE}. 
The main problem with applying weighted network analysis methods to aggregated data is that it assumes, explicitly or otherwise, that each aggregated set is an indivisible unit and that findings on the set apply to its constituent nodes. Such an assumption is termed an ecological fallacy, where findings on groups of individuals are extended to the individuals themselves~\cite{robinson2009ecological}.

The above-mentioned handling of aggregated networks is especially problematic for community detection methods. Few works applying community detection on aggregated datasets have attempted to confirm whether the community structure is conserved upon disaggregation. Instead, the relevance of these findings to the underlying individuals is either ignored or simply taken for granted~\cite{butts2009revisiting, Decuyper2018MeasuringTE}. 
Applying a disjoint community detection algorithm, where each aggregated set gets classified under one community, is problematic for at least three reasons. First, even in individual-level social networks, a categorical membership for a single community can be an oversimplification of an overlapping state of membership of several communities~\cite{yang2014overlapping,Kuppevelt2020CommunityMC}. The extent and composition of this overlap can be a distinguishing feature of nodes, reflecting the roles different nodes have in connecting the communities~\cite{Evans2009LineGL}.
Second, this oversimplification is exacerbated further in aggregated networks. When an entire group of individuals in an aggregated set is subject to one community classification, information is lost about the composition of its constituents' membership in other communities. Gandica et al. showed that detected disjoint community partitions over aggregated networks can be significantly sensitive to the partition and resolution of the aggregation, resulting in community classifications that are unstable across different aggregations~\cite{Decuyper2018MeasuringTE}. Sensitivity to the scale and shape of aggregation is a well-documented challenge, known as the Modifiable Areal Unit Problem (MAUP)~\cite{wong2004modifiable}. 
Third, it can be challenging to observe changes in community membership of nodes, as the disjoint classification of membership occludes small but meaningful changes~\cite{Xing2008ASM,peixoto2015inferring}. In aggregated networks, lacking the ability to distinguish these changes in membership hinders efforts to understand the evolution of membership composition of aggregated nodes and the community structure as a whole~\cite{rosvall2010mapping,cazabet2023challenges}.

Different methods have been proposed to address some of these concerns.  Probabilistic and latent-space models, such as the Mixed Membership Stochastic Block-Model (MMSBM)~\cite{Airoldi2007MixedMS} and overlapping SBM~\cite{peixoto2019bayesian}, successfully allow individual nodes to possess a vector of membership probabilities, reflecting overlapping communities. While some have adapted these approaches to aggregated relational data~\cite{jones2021scalable,ward2025bayesian}, they still face fundamental limitations when applied to coarse-grained networks. Specifically, models like MMSBM treat the aggregated set as a single, indivisible entity in a latent space, rather than as a collection of underlying constituents. Consequently, they lack a formal guarantee that the inferred mixed membership of an aggregate node corresponds to the sum of the mixed memberships of its unobserved constituent individuals.

To address this issue, we propose the Link Fraction Mixed Membership (LFMM). Rather than inferring a latent probability distribution, LFMM provides a strictly descriptive, algebraic metric for community composition in aggregated networks. The method defines the mixed membership of a node as the observed fraction of link volume connecting it to nodes within a given community partition. The primary advantage of LFMM over other mixed membership methods is its mathematical consistency across scales: because it is defined linearly, we prove that LFMM values computed on an aggregated network equal the sum of LFMM values computed on the disaggregated network. This method proves especially crucial in the case where one is interested in analyzing the mixed-membership structure of an aggregated network without access to the individual-level network. LFMM also stands out for being computable in a single matrix multiplication, being applicable to directed and weighted networks, and being compatible with any community detection algorithm. The method's sensitivity to aggregation is examined through numerical experiments on synthetic benchmark networks.

Then we utilize LFMM to investigate the community structure and evolution in a real-world aggregated population-scale social network of the Netherlands. The dataset is a register-based social network of all 17 million residents and the different types of affiliations (family, work, and school) connecting them, over 13 years~\cite{bokanyi2023anatomy,vanderlaan2022person}. Our mixed membership analyses focus on the aggregated forms of the network, where residents are grouped based on their residential addresses within approximately 3000 neighborhoods and 400 municipalities. While previous studies have identified disjoint, space-independent communities in a static Dutch social network of the Netherlands~\cite{Menyhert2024ConnectivityAC}, this work uses a mixed membership approach to investigate community diversity and yearly evolution over a decade. When applying LFMM, we find that mixed membership is heterogeneously distributed but strongly influenced by geospatial patterns. When accounting for the spatial factor using a gravity null model, we find that highly urban regions exhibit diverse connections across communities. Finally, we uncover significant longitudinal changes in the community structure and diversity of different regions. 

The remainder of this paper is organized as follows. Section~\ref{sec:Methodology} formalizes the Link Fraction Mixed Membership (LFMM) method, proves its consistency over aggregation, and defines the metrics used to quantify community diversity and statistical significance. In Section~\ref{sec:Results}, we present the population-scale social network of the Netherlands and apply LFMM to this dataset, revealing a strong correlation between urbanness and community diversity and capturing the evolution of the community structure. Finally, Section~\ref{sec:discussion} discusses the implications of the method and findings, and outlines directions for future research.

\section{Methodology and Validation}
\label{sec:Methodology}
In this section, we present the Link Fraction Mixed Membership (LFMM) method used to uncover community diversity in aggregated networks.
The analytical workflow employed in this study consists of two stages:
\begin{enumerate}
    \item \textbf{LFMM Computation:} LFMM requires applying an arbitrary community detection method to obtain a disjoint partition. It then evaluates mixed membership vectors for each aggregated node by computing the fraction of links connecting that node to each community (Section \ref{sec:LFMM}).
    \item \textbf{Diversity and Significance Analysis:} We quantify the heterogeneity of these membership vectors using a community diversity index and evaluate their statistical significance against a null model (Section \ref{sec:diversity}).
\end{enumerate}

\subsection{Link Fraction Mixed Membership}
\label{sec:LFMM}

The conceptual foundation of LFMM rests on a link-centric perspective of network structure, where a node's identity is defined by the distribution of its interactions. By defining mixed membership as the fraction of link volume connecting a node to a community, LFMM captures the association of a node or aggregate set with a community. Consequently, the method can be interpreted as a single-step diffusion process, representing the probability that a random walker starting at a node or set of nodes will land within a specific community. This approach draws upon the intuition that the node's role in networks is determined by its connectivity with the various communities, as also proposed in link clustering methods~\cite{Ahn2009LinkCR,cho2014mixed}.  

The LFMM method requires an initial disjoint partition of the aggregated network. For this study, we employ the Leiden algorithm~\cite{traag2019louvain} to optimize the Reichardt and Bornholdt’s Potts model~\cite{reichardt2006statistical}. This algorithm is chosen for its theoretical robustness under aggregation~\cite{Decuyper2018MeasuringTE}.

\subsubsection{LFMM formulation} 

We assume there is a hidden disaggregated weighted and undirected network $G=(V,E,W)$, represented by a symmetric adjacency matrix $\mathbf W$ with nonnegative weights $w_{ij}$ between nodes $i, j \in V$. Instead of $G$, we are given a graph $G'$, obtained by partitioning the nodes of $G$ into $n$ disjoint \textit{aggregation sets} $S_1,\dots,S_n$. The aggregated graph has $n$ nodes, and its weights $w'_{xy}$ are defined as the sum of weights in $G$ between nodes in the corresponding aggregation sets $S_x$ and $S_y$:
\begin{equation}
    w'_{xy}=\sum_{i\in S_x}\sum_{j\in S_y}w_{ij}.
    \label{eq:lfmm_aggregation}
\end{equation}
For self-loops, $w'_{xx}$ is the total weight of half-edges within set $S_x$. 
We denote the disjoint communities of the aggregated network by $C'_1,\dots,C'_r$. Each individual node is assigned the community label of its aggregation set, giving the corresponding individual-level communities $C_k=\bigcup_{x\in C'_k}S_x$.

\begin{figure}[h!]
    \centering
    \begin{subfigure}[t]{0.3\textwidth}
        \centering
        \includegraphics[width=\linewidth]{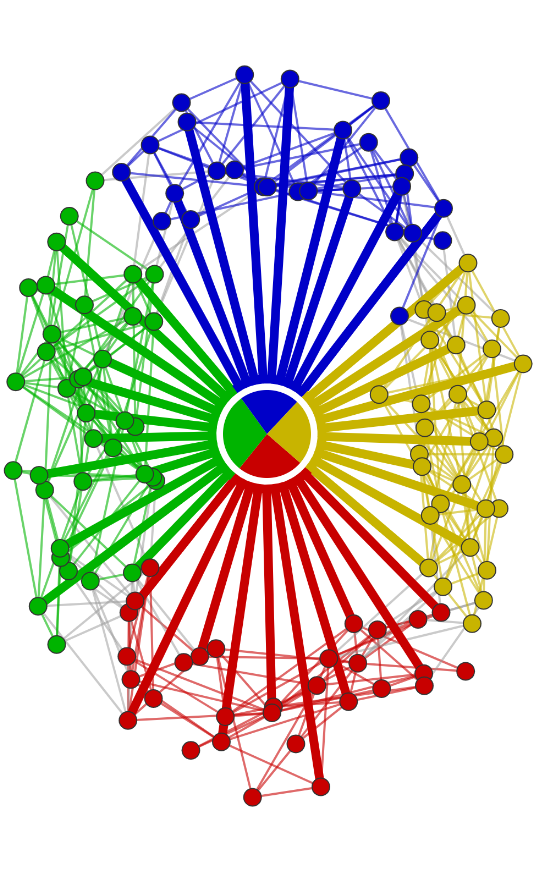}
        \caption{}
        \label{fig:lfmm_schematic_overview}
    \end{subfigure}
    \hspace{-3em}
    \begin{subfigure}[t]{0.74\textwidth}
        \centering
        \begin{tikzpicture}[scale=0.7, every node/.style={transform shape, align=center},
            annotation/.style={font=\small},
            arrow/.style={->, >=Stealth, thick},
            redarrow/.style={->, >=Stealth, thick, red}]
            \node[anchor=north] (top_img) at (5.,-0.5) {
                \includegraphics[height=0.22\textheight]{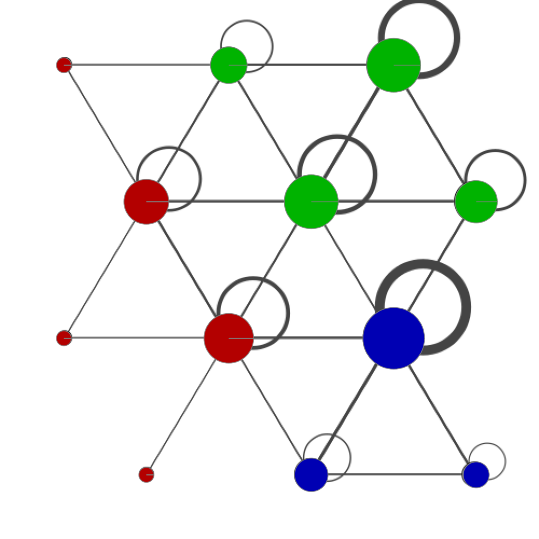}
            };
            \node[annotation, below=0.0cm of top_img] (lift_label) {\huge{$\Downarrow$} \normalsize \textbf{Disaggregate}};
            \node[below=0.0cm of lift_label] (bottom_img) {
                \includegraphics[height=0.22\textheight]{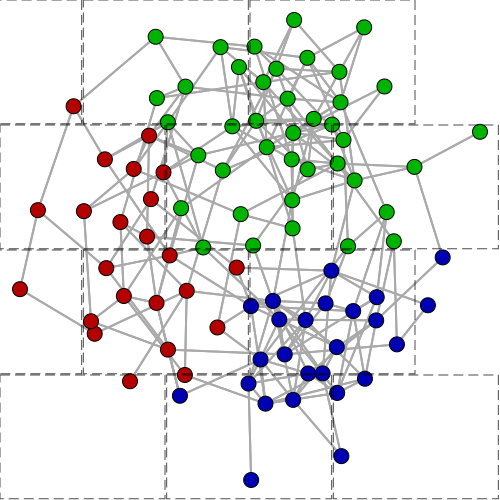}
            };
            \node[anchor=center, red] (arrow_top) at (8.5, -3) {\normalsize \textbf{LFMM} \\ \huge{$\Rightarrow$}};
            \node[anchor=center] (arrow_bottom) at (8.5, -8.5) {\normalsize \textbf{LFMM}  \\ \huge{$\Rightarrow$}};
            \node[anchor=north] (right_top) at (12, -0.5) {
                \includegraphics[height=0.22\textheight]{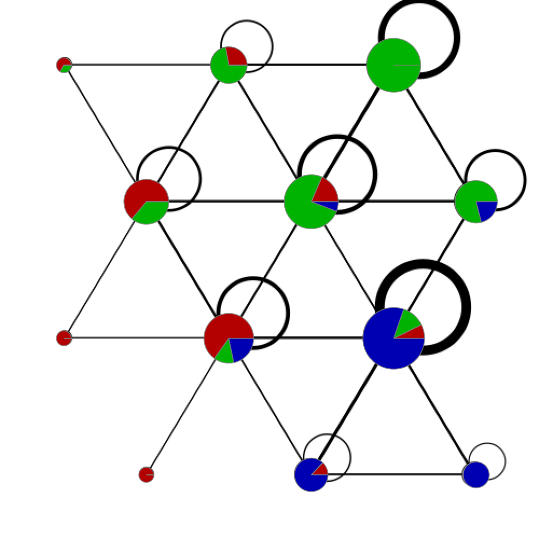}
            };
            \node[annotation, below=0.0cm of right_top] (sum_label) {\huge{$\Uparrow$} \normalsize \textbf{Aggregate}};
            \node[below=0.0cm of sum_label] (right_bottom) {
                \includegraphics[height=0.22\textheight]{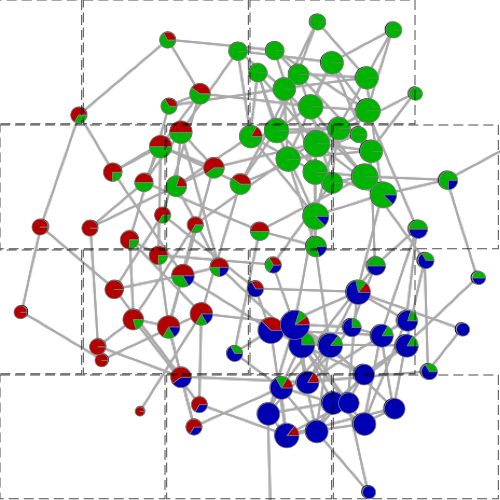}
            };
        \end{tikzpicture}
        \caption{}
        \label{fig:lfmm_schematic_paths}
    \end{subfigure}
    \caption{\textit{Link Fraction Mixed Membership (LFMM) method and its consistency under aggregation. } (a) Computing the link fraction of a node as fraction of connections to nodes within different community partitions. (b) Starting from an aggregated network with colored detected communities (top left): LFMM can be applied directly (red path) and be guaranteed to be equivalently computed by disaggregating communities to the individual-level, computing LFMM, and aggregating back (black path).  The dotted rectangles denote the aggregation partitioning of the individual-level network.
    }
    \label{fig:lfmm_schematic}
\end{figure}
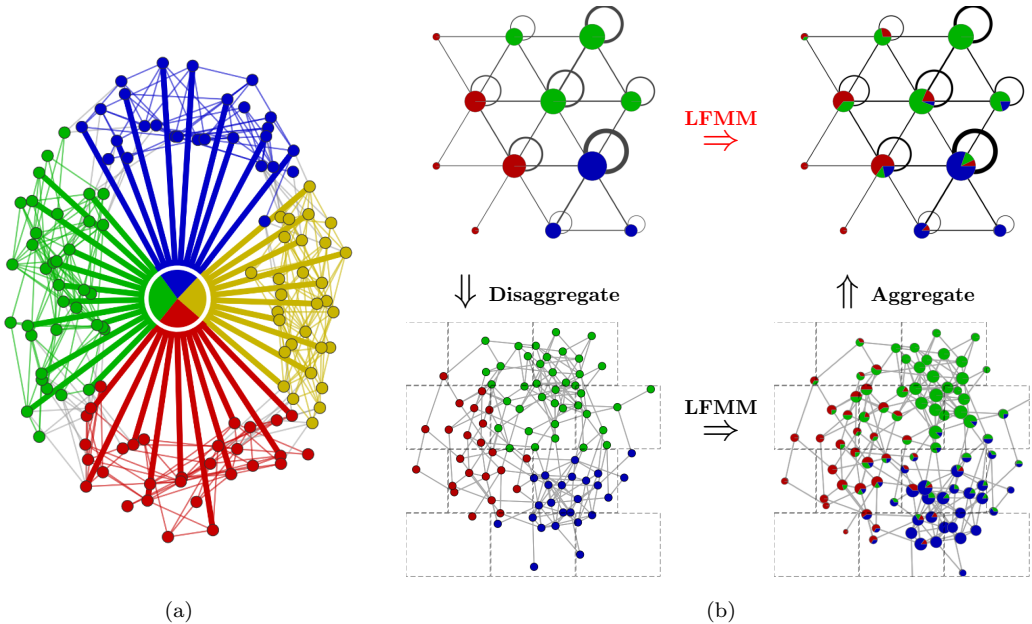

{
We now introduce the mixed membership vector $M$. For a node $i$ in the original graph, the unnormalized membership $M_i(k)$ of community $k$ and its normalized form $m_i(k)$ are defined as:
\begin{equation}
    M_i(k)=\sum_{j\in C_k}w_{ij},
    \qquad
    m_i(k)=\frac{M_i(k)}{s_i}.
    \label{eq:lfmm_individual}
\end{equation}
Here $s_i=\sum_j w_{ij}=\sum_k M_i(k)$ is the strength of node $i$.
The normalized mixed membership $m_i(k)$ is the link-weight fraction
of node $i$ towards nodes labeled under community $k$, as illustrated
in Figure~\ref{fig:lfmm_schematic_overview}.

Analogously, for an aggregate set $S_x$, we define the aggregate mixed membership $M'_x$ and its normalized form $m'_x$ using the aggregated weights:
\begin{equation}
    M'_x(k)=\sum_{y\in C'_k}w'_{xy},
    \qquad
    m'_x(k)=\frac{M'_x(k)}{s'_x}.
    \label{eq:lfmm_aggregate}
\end{equation}
Here $s'_x=\sum_y w'_{xy}=\sum_k M'_x(k)$ is the strength of the aggregate node, including internal connections. The unnormalized mixed membership matrix for the entire network can be calculated via a single matrix multiplication of the aggregated adjacency matrix and a community indicator matrix. The matrix formulation and an extension of the method as a diffusion process that accounts for higher-order connectivity are provided in Appendix~\ref{app:matrix_comp}.

\subsubsection{LFMM consistency under aggregation} 
\label{sec:consistency}

A key property of LFMM is that, due to the linearity of the formulation, it is consistent under aggregation. More specifically, for an aggregated set $S_x$, the sum of the unnormalized mixed membership vectors $M_i$ of its constituent nodes equals the mixed membership $M'_x$ computed directly on the aggregated set. Using the correspondence between the communities $C_k$ and $C'_k$ defined above, we obtain:
\begin{align}
    \sum_{i\in S_x}M_i(k)
    &=\sum_{i\in S_x}\sum_{j\in C_k}w_{ij} \nonumber\\
    &=\sum_{y\in C'_k}\left(\sum_{i\in S_x}\sum_{j\in S_y}w_{ij}\right)
      \nonumber\\
    &=\sum_{y\in C'_k}w'_{xy}=M'_x(k).
    \label{eq: conservation}
\end{align}
The same property holds for node strength, since the strength of an aggregated set is the sum of the strengths of its constituent nodes:
\begin{equation}
    s'_x=\sum_y w'_{xy}
        =\sum_{i\in S_x}\sum_j w_{ij}
        =\sum_{i\in S_x}s_i.
    \label{eq:lfmm_strength}
\end{equation}
Combining these two relationships, the normalized mixed membership
$m'_x(k)$ can be expressed as:
\begin{equation}
    m'_x(k)
    =\frac{M'_x(k)}{s'_x}
    =\frac{\sum_{i\in S_x}M_i(k)}{\sum_{i\in S_x}s_i}
    =\frac{\sum_{\substack{i\in S_x}}s_i m_i(k)}
           {\sum_{i\in S_x}s_i}.
    \label{eq:lfmm_strength_weighted}
\end{equation}
Here we assume $s'_x>0$, to prevent division by zero. Nodes with zero strength make no contribution to the aggregate membership. Thus, the normalized LFMM of an aggregated set equals the strength-weighted average of the normalized LFMM values of its constituent nodes. }

This conservation is illustrated in Figure~\ref{fig:lfmm_schematic_paths}, which contrasts two computational pathways to arrive at the mixed membership values in the aggregated network. Starting from a disjoint community partition of the aggregated network, the first path (red arrow) represents the direct application of LFMM. The second path (black arrows) represents a theoretical process where the community labels are assigned to the constituent individual nodes, LFMM is computed for every node, and the results are aggregated back. These two pathways are mathematically equivalent when the unnormalized vectors are summed, or when the normalized vectors are averaged in proportion to node strength. Both pathways use the same community labels. Consequently, this conservation does not imply statistical recovery of an unknown individual-level ground truth, or agreement between community partitions detected independently at different scales.

A caveat for this definition is that it is edge-centric, meaning that nodes with higher degree or strength play a larger role in the mixed membership of their aggregated set. This remains true for the normalized form: $m'_x$ describes the fraction of link volume connecting the set to each community, rather than the fraction of individuals belonging to that community. An unweighted average of the constituent $m_i$ values therefore need not equal $m'_x$, although the two coincide when all constituent nodes have equal positive strength. Differences between these averages reflect the choice of weighting, rather than a failure of the consistency property.

\subsubsection{Synthetic Networks and Benchmarks}
\label{sec:data_synthetic}

To validate and analyze the consistency of the LFMM method and its robustness against aggregation, we generate synthetic individual-level networks using the Stochastic Block Model (SBM)~\cite{abbe2018community}. Let $G$ be an unweighted, undirected graph with $N=1,000$ nodes, partitioned into $r=2$ equally sized ground-truth communities. The edge generation is governed by an expected average degree $\langle k \rangle = 20$ and an \textit{affinity parameter} $\mu \in [0, 1]$, defined as the fraction of a node's expected edges that connect to the opposing community. To simulate the coarse-graining process, we project $G$ into an aggregated network $G'$ consisting of $n=50$ aggregate sets $\textbf{S}$. We control the alignment between the true communities and the aggregate boundaries using an \textit{aggregation mixing parameter} $m \in [0, 1]$. For a node belonging to ground-truth community $c$, with probability $1-m$, it is assigned to an aggregate set exclusively reserved for community $c$; with probability $m$, it is assigned uniformly at random to any of the 50 aggregate sets.

\begin{figure}[ht!]
    \centering
    \begin{subfigure}[t]{0.45\textwidth}
        \centering
        \includegraphics[width=\linewidth]{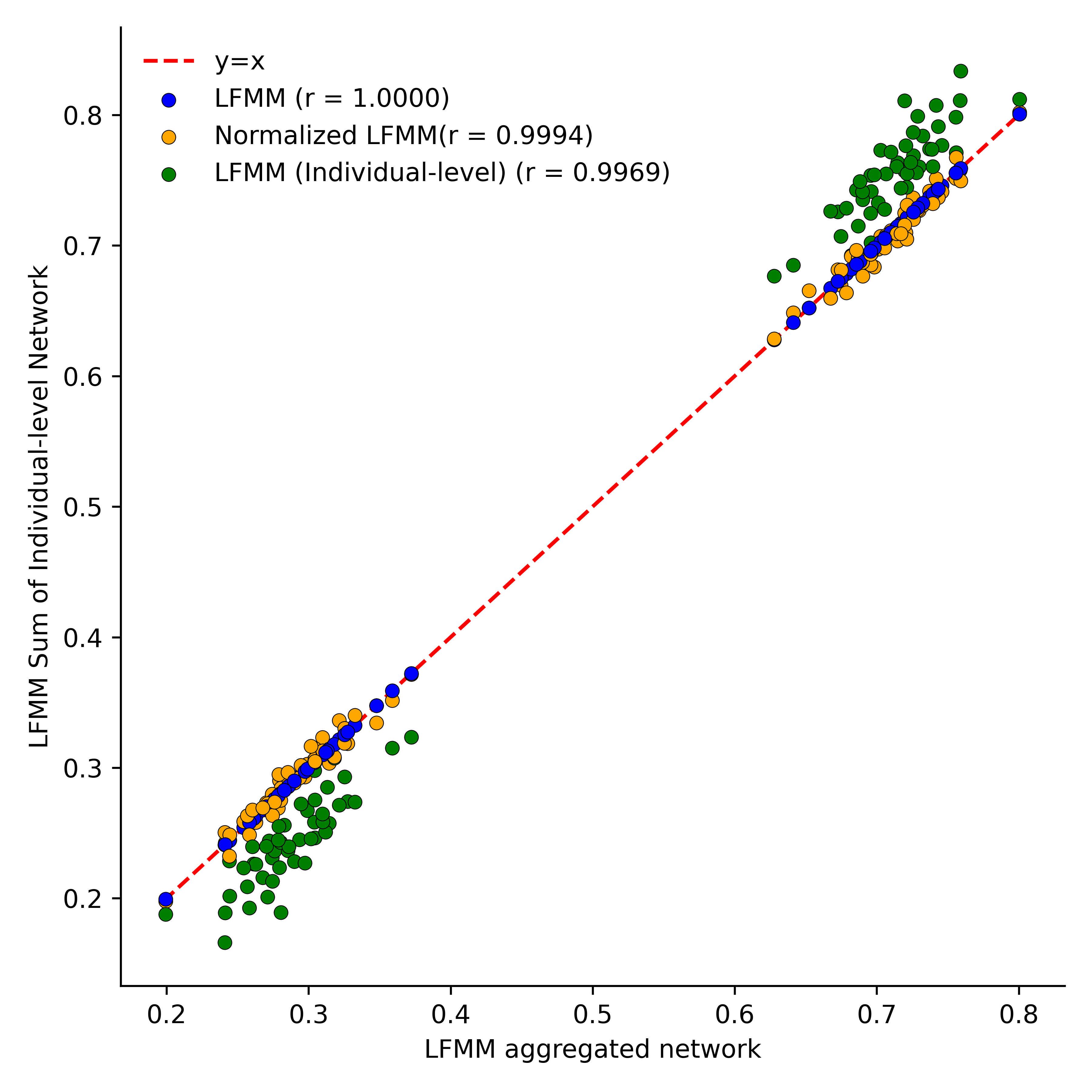}
        \caption{}
        \label{fig:synthetic_consistency}
    \end{subfigure}
    \hfill
    \begin{subfigure}[t]{0.5\textwidth}
        \centering
        \includegraphics[width=\linewidth]{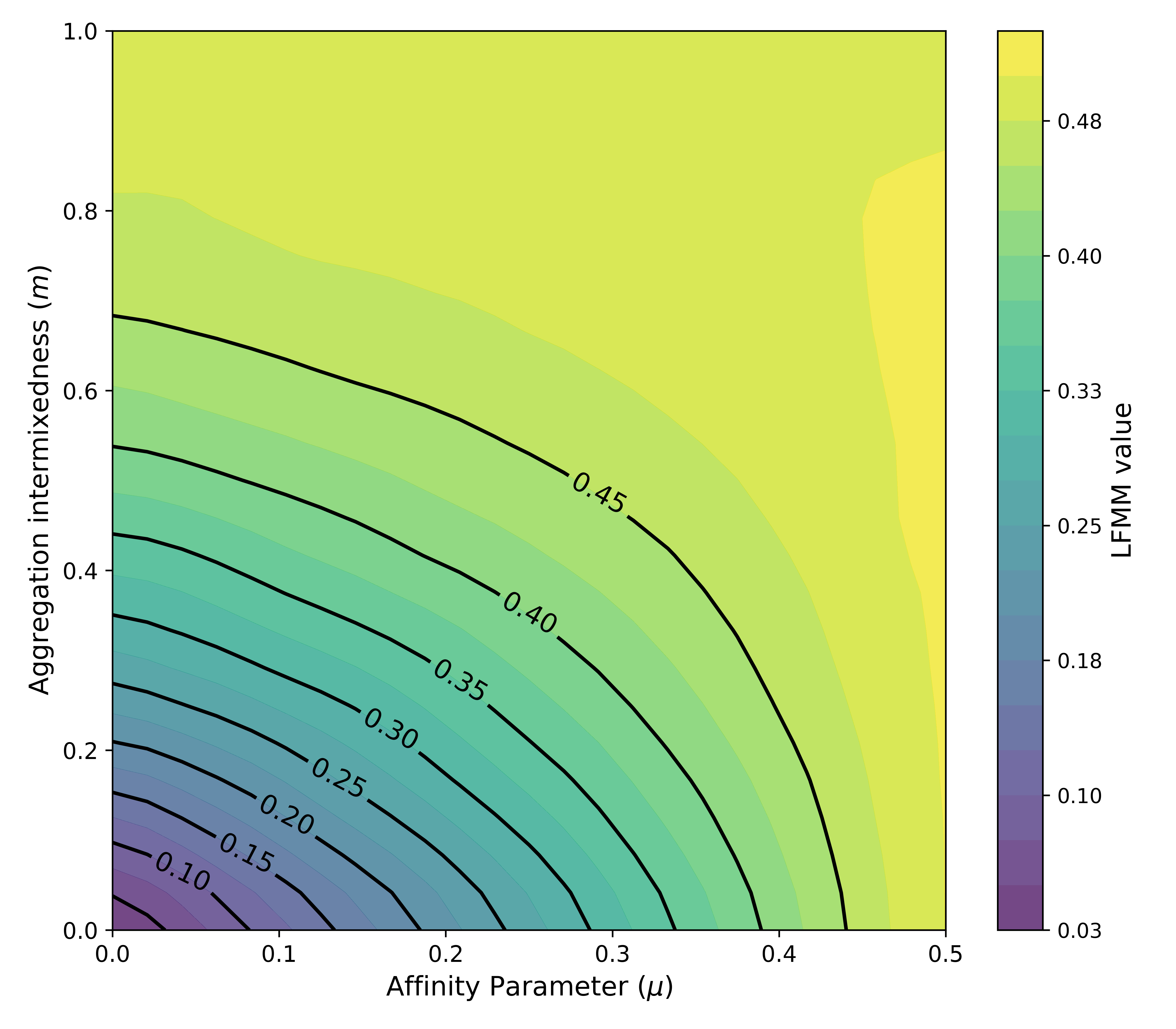}
        \caption{}
        \label{fig:synthetic_heatmap}
    \end{subfigure}
    \caption{\textit{LFMM consistency under aggregation and community affinity.} \textbf{(a)}~Sum of mixed memberships computed on the synthetic individual-level network ($m=0.2$) versus the mixed memberships computed directly on the aggregated network. \textbf{Blue:} LFMM values ($M$) following disaggregation then re-aggregation. \textbf{Orange:} Normalized LFMM values ($m$). \textbf{Green:} LFMM of community detection performed at the individual-level. \textbf{(b)} Isoline contour plot of mean LFMM value for different values of affinity ($\mu$) and aggregation intermixedness $m$.}
    \label{fig:synthetic_validation}
\end{figure}

We perform three comparisons to evaluate the consistency of LFMM after applying community detection using the Leiden Algorithm~\cite{traag2019louvain}, visualizing the correlation between values computed directly on the aggregated network $G'$ and those derived from the individual-level network $G$. First, we validate the conservation property (eq. \ref{eq: conservation}) by comparing the raw LFMM vector $M'$ (as defined in Section \ref{sec:LFMM}) computed on $G'$ against the sum of individual vectors $M$ computed on $G$ (using the community partition from the aggregate network). Second, we repeat this comparison for the sum of normalized vectors $m$ versus $m'$ to assess deviations caused by the normalization of mixed membership vector. Finally, we compare the aggregate LFMM results against a "ground truth" scenario where community detection is performed on the individual-level graph $G$ rather than $G'$. 

The results are displayed in Figure~\ref{fig:synthetic_consistency}. The raw LFMM values (blue) exhibit a perfect correlation ($r=1.0$), empirically confirming that the method is mathematically consistent under aggregation. The normalized values (orange) show a high but imperfect correlation ($r \approx 0.999$), as the normalization by node strength does not scale linearly with aggregation. The comparison with the individual-level community detection (green) demonstrates that LFMM applied to aggregated data ($r \approx 0.997$) also serves as a reasonable approximation of the underlying micro-scale community structure, though it shows a bias of over-estimating the minority membership in aggregated nodes. Practically, this bias occurs when attempting to capture the unknown community structure at the individual-level, as suboptimal partitioning creates a deviation between the aggregated and individual scales.

When evaluating the mean values of LFMM for synthetic networks of various affinity parameters and aggregation mixing, we find that mixed membership highly correlates with either affinity or mixing when the other factor is absent (see Figure \ref{fig:synthetic_heatmap}). However, when combined, there is a nearly symmetric effect of both factors on the mixed membership values. This happens because high aggregation mixing (aggregated sets that contain a heterogeneous mix of members of both communities) results in a link fraction similar to a well-aggregated network (homogeneous sets) that has high affinity. In other words, under the assumption of a disjoint community structure in the individual layer, LFMM does not distinguish between an aggregated set that is heterophilically connected and one that is heterogeneously mixed. Consequently, distinguishing which is the source of high mixed membership can only be achieved through inspecting the alignment of community membership at the individual level with the aggregated sets. 
{
When the individual-level network is available, this alignment can be assessed using the aggregability index $\eta$ proposed in \cite{Decuyper2018MeasuringTE}, which measures how much information about an individual's community membership is provided by their aggregation set. As  LFMM is designed for aggregated networks, its description of aggregate-level connectivity and its algebraic consistency hold regardless of the value of $\eta$.} 

\subsection{Community diversity and statistical significance}
\label{sec:diversity}

For the purposes of empirically testing LFMM on a large-scale social network in Section 3, we describe two necessary metrics for evaluating membership diversity and statistical significance.

First, to measure the overall diversity of community memberships within a given aggregate set, we employ the Gini-Simpson Index (GSI)~\cite{Jost2006}. Given the normalized mixed membership vector $\vec{m}_i$ for an aggregate set $i$, the GSI is calculated as:
\begin{equation}
    GSI(\vec{m}_i) = 1 - \sum_{j=1}^{r} m_{i}(j)^2.
\end{equation}
Here $r$ is the total number of communities. In the context of an aggregate set, GSI represents the probability that two individuals, drawn at random from the aggregate set $S$, belong to different communities. The index ranges from 0 (single community) to a theoretical maximum of $1 - 1/r$ (uniform distribution). A high GSI value thus corresponds to a high degree of local co-existence between members of different communities. While this diversity metric is directly proportional to the total minority membership fraction, it distinguishes two high minority fractions based on how heterogeneous the membership distribution is.

Second, to isolate the component of diversity that is not explained by geographic proximity, we compute the statistical significance of the GSI diversity, via the $z$-score, as compared to a gravity null model~\cite{prieto2018gravity}. This metric compares the empirically observed GSI with the expected GSI mean and variance derived from the gravity null model. The $z$-score $z_i$ for aggregate set $i$ is defined as:
\begin{equation}
    z_i = \frac{GSI_i - \mu_i}{\sigma_i}.
\end{equation}
Here, $\mu$ and $\sigma$ are the gravity null model mean and standard deviation for the set. A z-score value near zero indicates that the observed diversity value is expected based on the region's relative geographic location and population.

\section{Community mixed membership in the Dutch social network}
\label{sec:Results}

In this section, we present the aggregated social network of the Netherlands and empirical results obtained by applying the proposed Link Fraction Mixed Membership method. We provide two sets of analyses. The first pertains to the computation and analysis of mixed membership values, in comparison with disjoint community partitions, and tracking community evolution over time. The second evaluates the diversity of community memberships, distinguishing between spatially-driven and socially-driven heterogeneity. In the latter, we utilize a spatial null model to identify significant diversity patterns and link them to the level of urbanization in different regions.

\subsection{Population-scale Social Network of the Netherlands}
\label{sec:data_netherlands}

We utilize the register-based population-scale social network of the Netherlands. This dataset is constructed from yearly administrative registers covering the entire population of the country (approximately 17 million residents and 1 billion edges each year). The network captures formal social ties defined by government records, including family relationships (first- and second-degree relatives, partners), school affiliations (primary, secondary, vocational school, and university year groups), household connections, next-door neighbors, and work relationships (colleagues) \cite{vanderlaan2022person, bokanyi2023anatomy}. By combining these layers, the network represents the "social opportunity structure" of the population \cite{soler2024contacts,bokanyi2023anatomy}.

For the purposes of this study, we focus on two spatial aggregations of the above individual-level networks, based on people's residential addresses within \textit{neighborhoods}, which are in turn part of \textit{municipalities}. This results in two undirected weighted networks per year: a neighborhood-aggregated network and a municipality-aggregated network, where edge weights represent the sum of all types of social ties between residents of any two regions. Since household and next-door neighbor connections almost never cross neighborhood boundaries by definition, we omit them from our analysis. For family, school, and work connections, our aggregation preserves the internal connectivity within each administrative unit (i.e., number of half-edges between residents of the same neighborhood) as self-loops. Access to this value is necessary for the conservation property of the LFMM method.

The network is available for each year from 2009 to 2021, allowing for the analysis of the evolution of community structure over time. For analysis results of a single year, we focus on the 2021 snapshot of the network (0.8 billion edges over 3218 neighborhoods or 352 municipalities). 
Administrative boundaries defining these aggregation sets are not static; municipal re-divisions, mergers, and border adjustments occur often, which consequently alter the composition of the aggregation sets across snapshots.

\subsection{Mixed membership in the Dutch social network}

\begin{figure}[t]
    \centering
    \begin{subfigure}[b]{0.43\textwidth}
        \includegraphics[width=\linewidth]{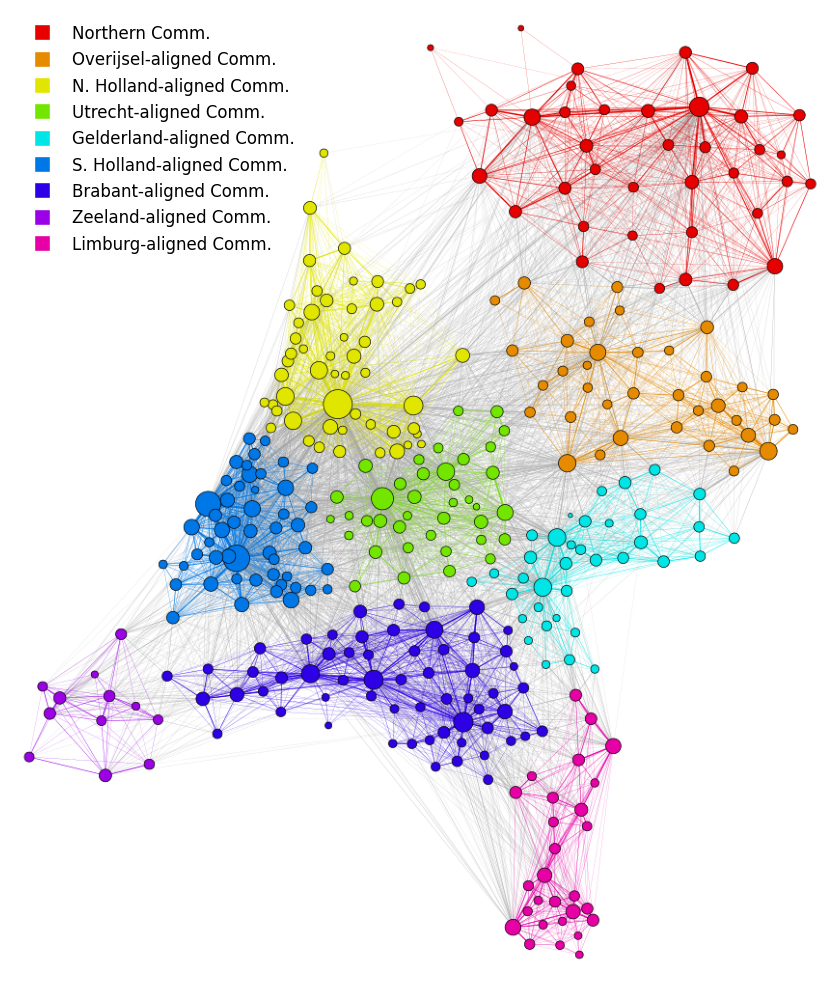}
        \caption{}
        \label{fig:community_a}
    \end{subfigure}
    \hfill
    \begin{subfigure}[b]{0.55\textwidth}
        \includegraphics[width=\linewidth]{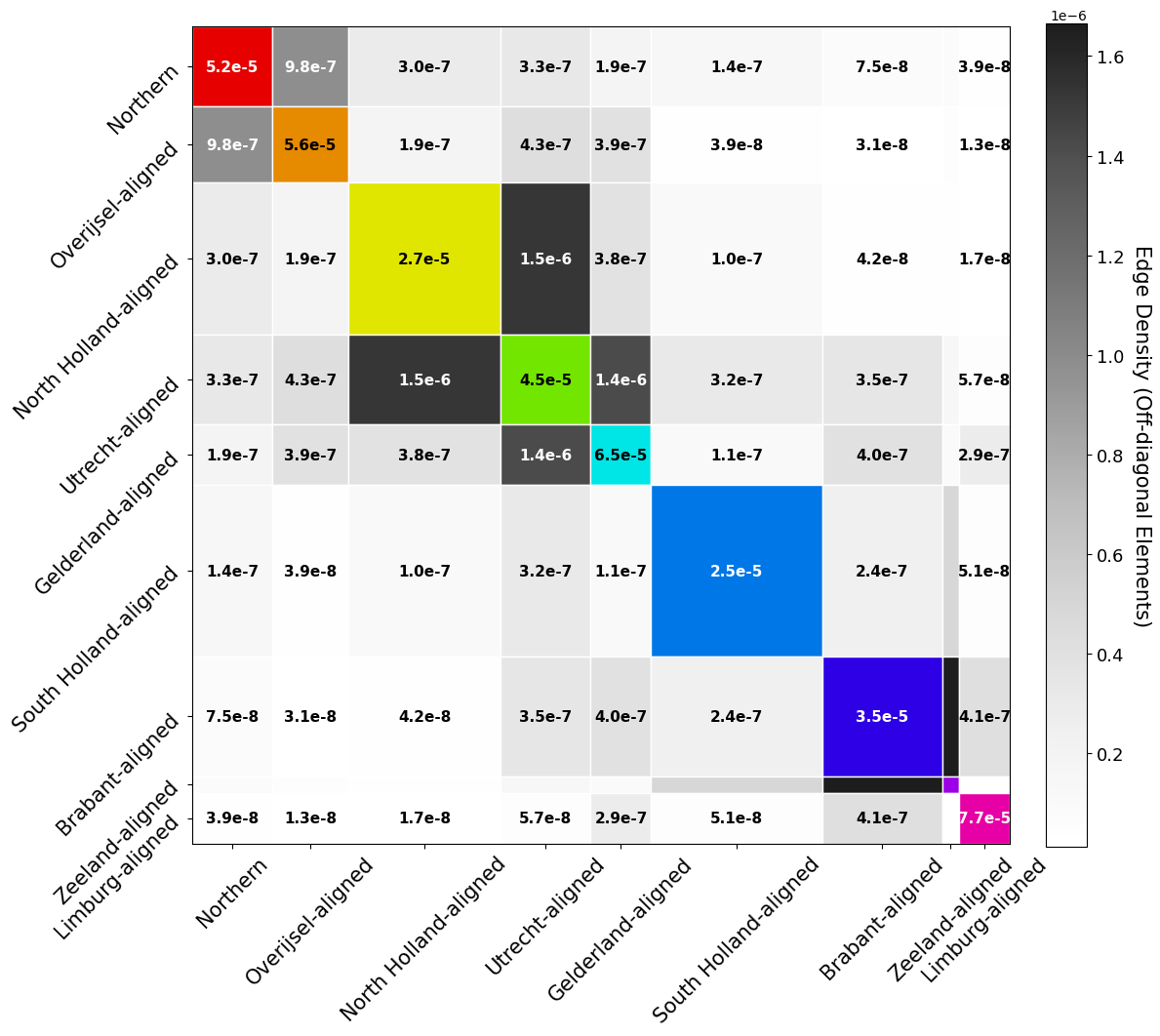}
        \caption{}
        \label{fig:community_b}
    \end{subfigure}
    \caption{\textit{Disjoint community structure of the population-scale network of the Netherlands} \textbf{(a)} Community partition of the municipality-aggregated network. Node sizes are proportional to population, colors represent community membership. Communities are named after the provinces they most closely match. \textbf{(b)} Community-aggregated adjacency matrix showing the density of connections within (diagonal) and between (off-diagonal) communities. Block sizes are proportional to community population. The matrix is ordered to minimize off-diagonal density values.}
\label{fig:community}
\end{figure}

We first identify the large-scale community structure using a disjoint community detection on the municipality-aggregated network of the population-scale social network of the Netherlands. The resulting partition (Figure~\ref{fig:community_a}) and the community-aggregated adjacency matrix (Figure~\ref{fig:community_b}) reveal a strong spatial embedding, with communities forming geographically contiguous territories that closely align with provincial administrative borders, similar to other findings in the literature~\cite{robiglio2025multiscale, Menyhert2024ConnectivityAC,kallus2015spatial}. Communities were named after the province or administrative region with which they had the most nodes in common (e.g., the green community is labeled "Utrecht-aligned community" as it is centered around the province).

\begin{figure}[t!]
    \centering
    \begin{minipage}[c]{0.7\textwidth}
        \centering
        \includegraphics[width=\linewidth]{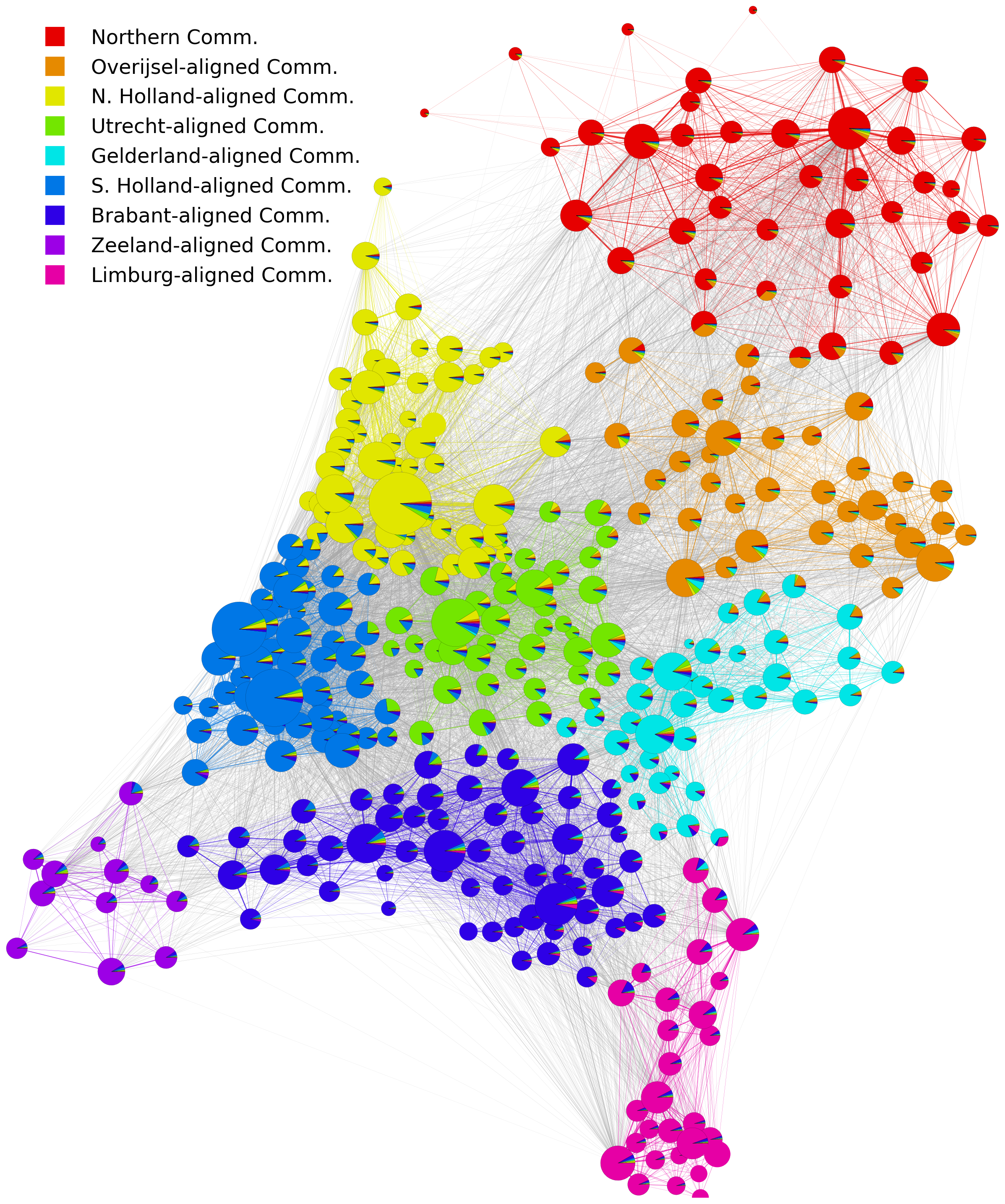}
        \subcaption{}
        \label{fig:pies_a}
    \end{minipage}
    \hfill
    \begin{minipage}[c]{0.29\textwidth}
        \centering
        \includegraphics[width=\linewidth]{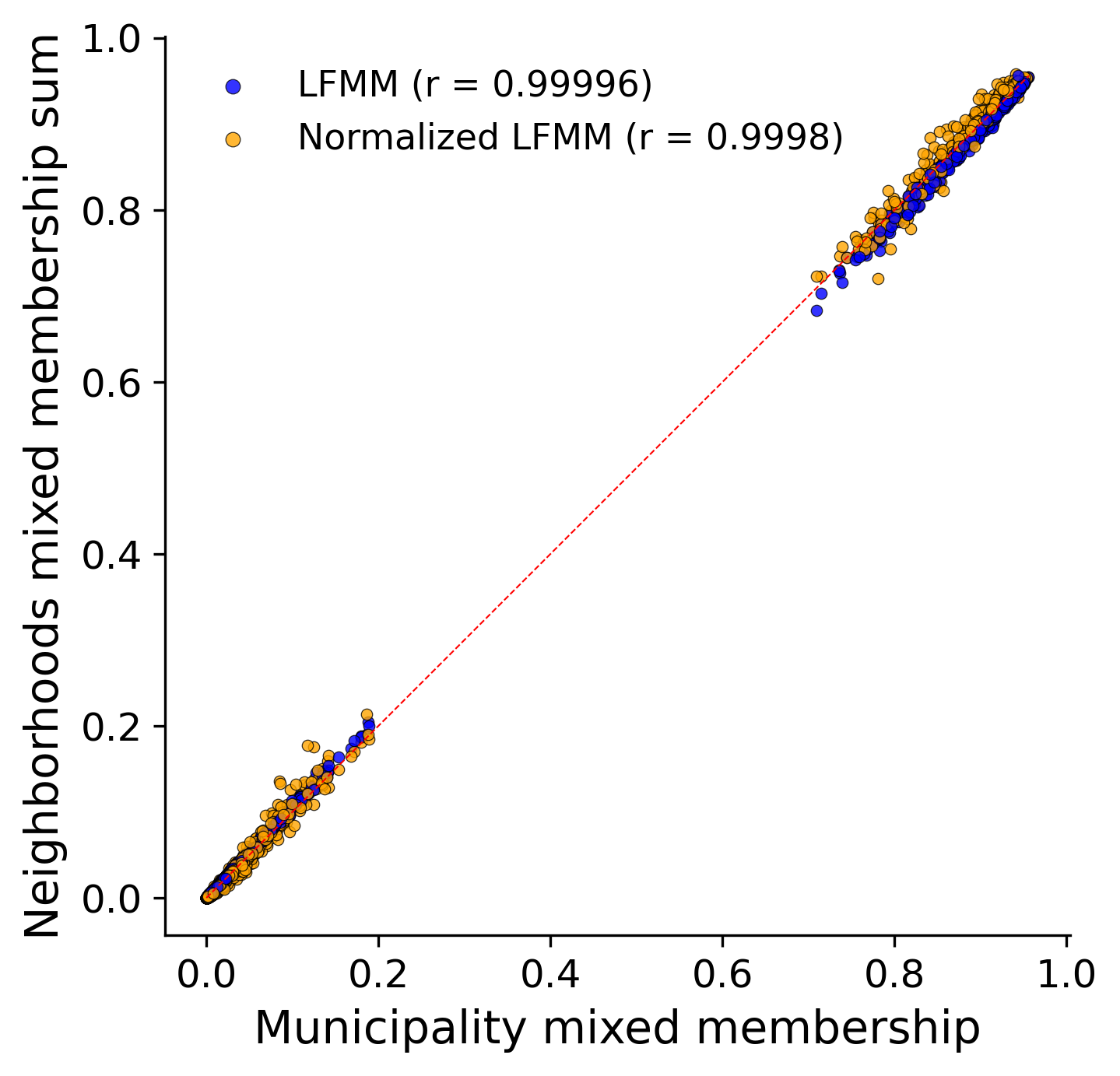}
        \subcaption{}
        \label{fig:pies_b}
        \vspace{1em}
        
        \includegraphics[width=\linewidth]{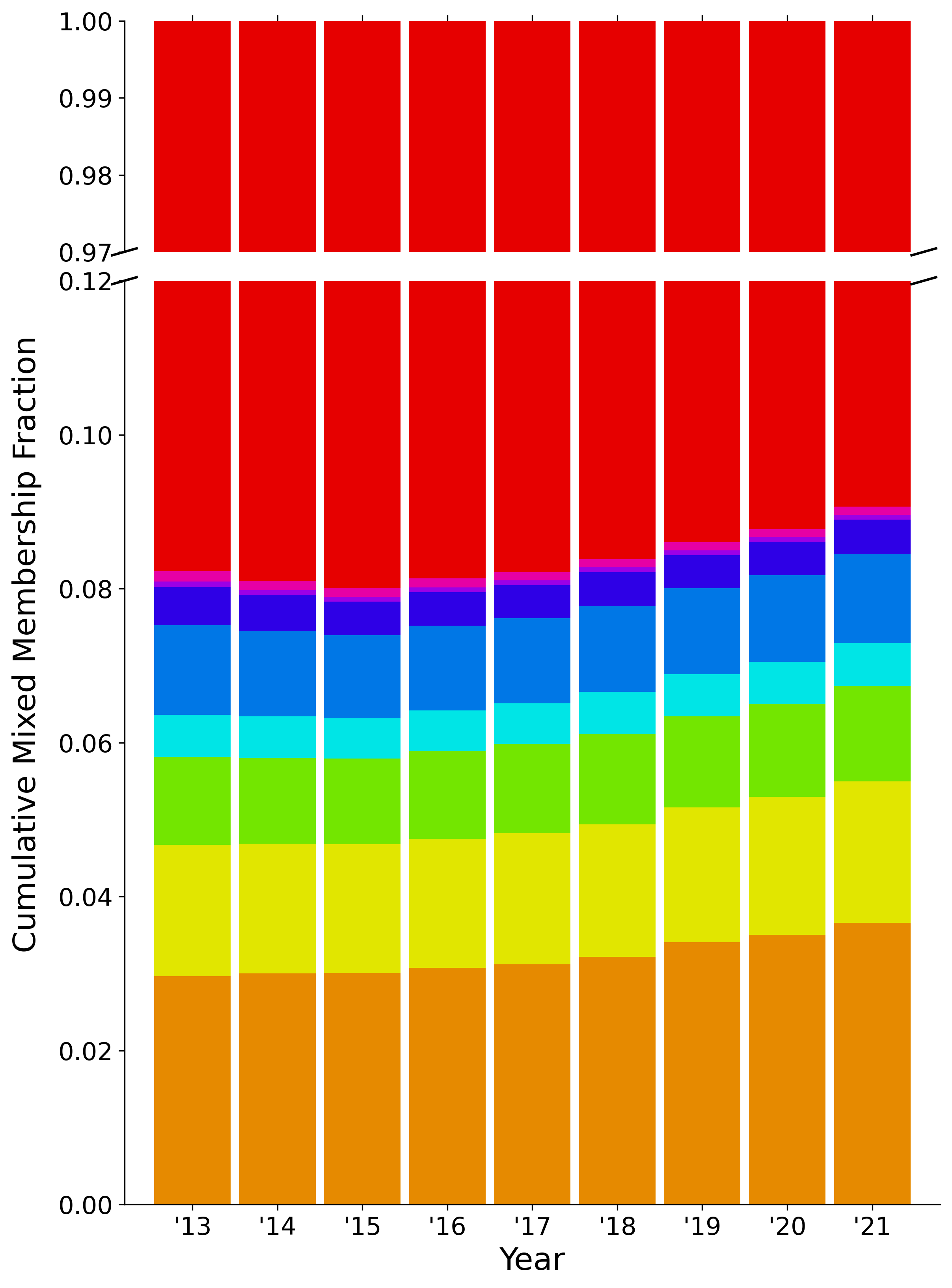}
        \subcaption{}
        \label{fig:pies_c}
        \hfill
    \end{minipage}
    \caption{\small \textit{Mixed membership community composition and analysis}. \textbf{(a)} The Dutch municipality-aggregated network. Each node is represented by a pie chart showing the distribution of its mixed membership vector, sized proportional to population. Colors correspond to the communities identified in Figure~\ref{fig:community}. \textbf{(b)} Comparison of LFMM values computed directly on the municipality-aggregated network (vertical axis) versus the sum of LFMM values computed on the neighborhood network (horizontal axis). \textbf{(c)} Temporal evolution over a decade of the total mixed membership mass within the northern (red) community, broken down by target community affiliation.}
    \label{fig:pies}
\end{figure}

{To assess whether the geographical aggregation is aligned with the underlying community structure, we applied the same community detection algorithm to the unweighted individual-level network for 2021 and computed the aggregability index $\eta$~\cite{Decuyper2018MeasuringTE}. As outlined at the end of Section~\ref{sec:data_synthetic}, this index measures how much information about an individual's community membership is provided by their aggregation set. We find high aggregability at both the municipality ($\eta = 0.931$) and neighborhood ($\eta = 0.946$) levels, indicating a strong alignment between the administrative units and the detected individual-level communities. This alignment supports the relevance of the subsequent mixed membership and diversity analyses for the residents of the aggregated geographic regions.} 

To investigate the internal composition of these regions, we applied LFMM (as defined in Section \ref{sec:LFMM}) to obtain the mixed membership composition for each municipality (Figure~\ref{fig:pies}). The resulting membership distribution improves upon the limitations of the disjoint partition described above. While the disjoint algorithm enforces a sharp boundary between communities, the LFMM results show that municipalities situated on opposite sides of these detected borders exhibit similar mixed membership profiles. The mixed membership view also shows that spatial proximity drives a continuous transition of community influence, rather than discrete territories. However, certain deviations from spatial patterns can be observed, namely in a distinction between rural and urban areas. While rural municipalities are largely dominated by a single community membership, major urban centers such as Utrecht, Amsterdam, and Rotterdam display a noticeably more heterogeneous composition of non-local memberships, as can be seen in Figure \ref{fig:pies_a}. However, to rule out the possibility that this pattern is merely an artifact of the municipalities' size and geographic centrality, we validate this observation against a spatial null model in Section~\ref{sec:result diversity}. The mixed membership values for the 40 largest municipalities can be found in the Appendix Table~\ref{tab:top40_municipalities}.

After having investigated the method's consistency across different scales of aggregation on simulated data in Section \ref{sec:data_synthetic}, we now do so empirically on the Dutch network in Figure~\ref{fig:pies_b}. We observe that the mixed membership values computed directly on the coarser municipality network are nearly identical to the sum of the values computed on the finer neighborhood network (Pearson correlation $\rho > 0.999$). This confirms that LFMM is robust against the aggregation level of the network, preserving the total "mass" of community membership regardless of whether the network is analyzed at the neighborhood or municipal scale.

Finally, the method enables the analysis of gradual shifts in the community structure that are unobservable in the disjoint partitioning. Figure~\ref{fig:pies_c} tracks the evolution of total mixed membership within the northern community over a decade. We observe gradual changes in the community composition, such as the rising influence of the neighboring Overijssel-aligned (orange) community and the decline of connections to the southern provinces. This demonstrates the method's capacity to capture slow shifts in the mesoscale network structure over time not visible through traditional methods.

\subsection{Community diversity}
\label{sec:result diversity}

\begin{figure}[ht!]
    \centering
    \begin{subfigure}[t]{0.38\textwidth}
        \centering
        \includegraphics[width=\linewidth]{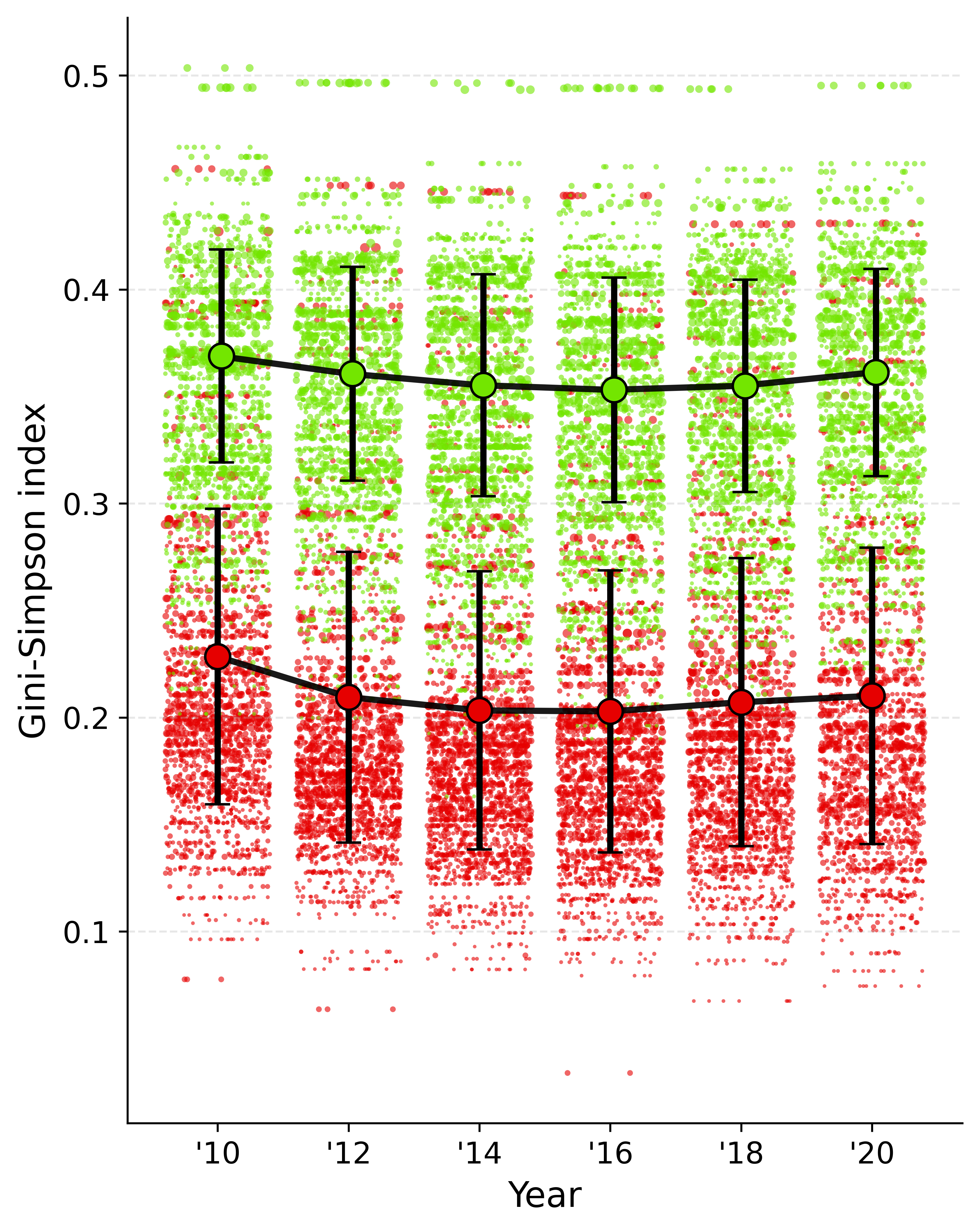}
        \caption{}
        \label{fig:gsi_temporal}
    \end{subfigure}
    \hfill
    \begin{subfigure}[t]{0.6\textwidth}
        \centering
        \includegraphics[width=\linewidth]{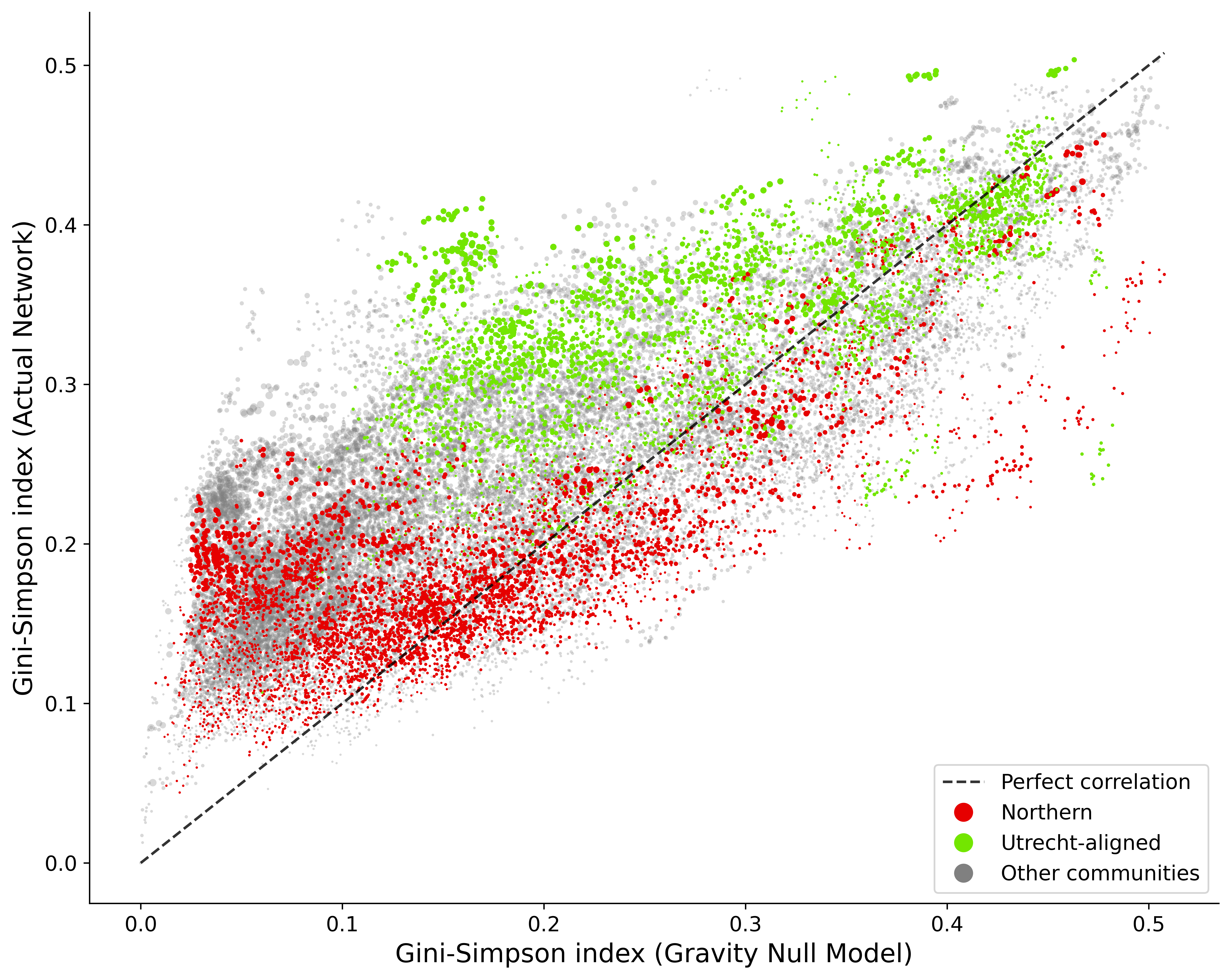}
        \caption{}
        \label{fig:gsi_gravity}
    \end{subfigure}
    \caption{\small \textit{Community diversity in the neighborhood-aggregated network over time.}  \textbf{(a)} Temporal evolution of the Gini-Simpson diversity index (GSI) diversity of neighborhoods in the Utrecht-aligned (green) and Northern (red) communities from 2010 to 2020. Points are neighborhoods and scaled by population size. \textbf{(b)} GSI for each neighborhood in the observed network (vertical axis) versus the GSI expected from the gravity null model (horizontal axis). Points near the dashed diagonal line have a diversity value that is expected by the null model. 
    }
    \label{fig:gsi_scatter}
\end{figure}

We quantify the heterogeneity of community composition within each municipality through employing the GSI diversity metric (as defined in Section~\ref{sec:diversity}) on the mixed membership values. First, in Figure~\ref{fig:gsi_temporal}, we inspect the evolution of diversity of neighborhoods in the Utrecht-aligned (green) and northern (red) communities over a decade. Significant changes in mean diversity were observed in both communities, with a significant drop from 2010 values that is partially recovered later on by the Utrecht-aligned community. Second, since we are interested in how the diversities of neighborhoods within different communities compare both to one another and to a spatial null model, we plot the observed diversity against the diversity expected from the gravity null model in Figure~\ref{fig:gsi_gravity}. 
{ Although the gravity null model does not explicitly control for node strength, it is not necessary for this particular dataset, since at this level of aggregation the average strength of residents is relatively uniform across municipalities and neighborhoods in the Dutch network.} 
A strong correlation is evident, confirming that the relative geographic location is a primary determinant of a region's diversity. For instance, centrally located communities like the Utrecht-aligned community exhibit high diversity in the null model and somewhat higher values in the observed network, simply due to their proximity to multiple other communities. Conversely, peripheral communities like the northern community show low expected diversity. However, many neighborhoods exhibit diversity scores significantly higher than what the gravity model predicts, suggesting the presence of social forces beyond spatial considerations. This deviation is not uniform across communities; for example, while neighborhoods in the northern community cluster at low expected diversity values, a subset of them diverges from the diagonal, indicating much higher actual diversity than their geography would predict.

\begin{figure}[ht!]
    \centering
    \begin{subfigure}[t]{0.44\textwidth}
        \centering
        \includegraphics[width=\linewidth]{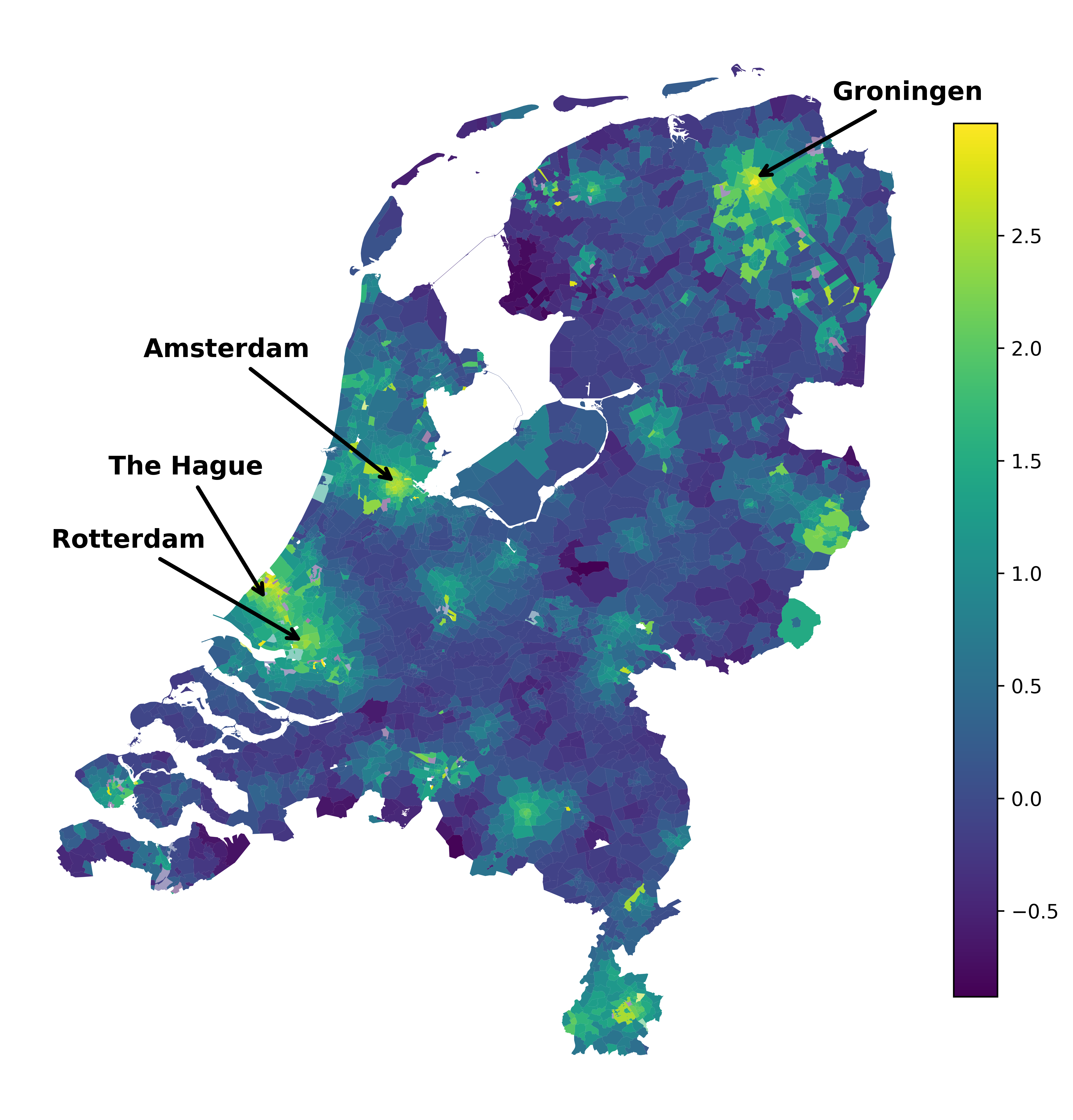}
        \caption{}
        \label{fig:sidi_urban_a}
    \end{subfigure}
    \hfill
    \begin{subfigure}[t]{0.55\textwidth}
        \centering
        \includegraphics[width=\linewidth]{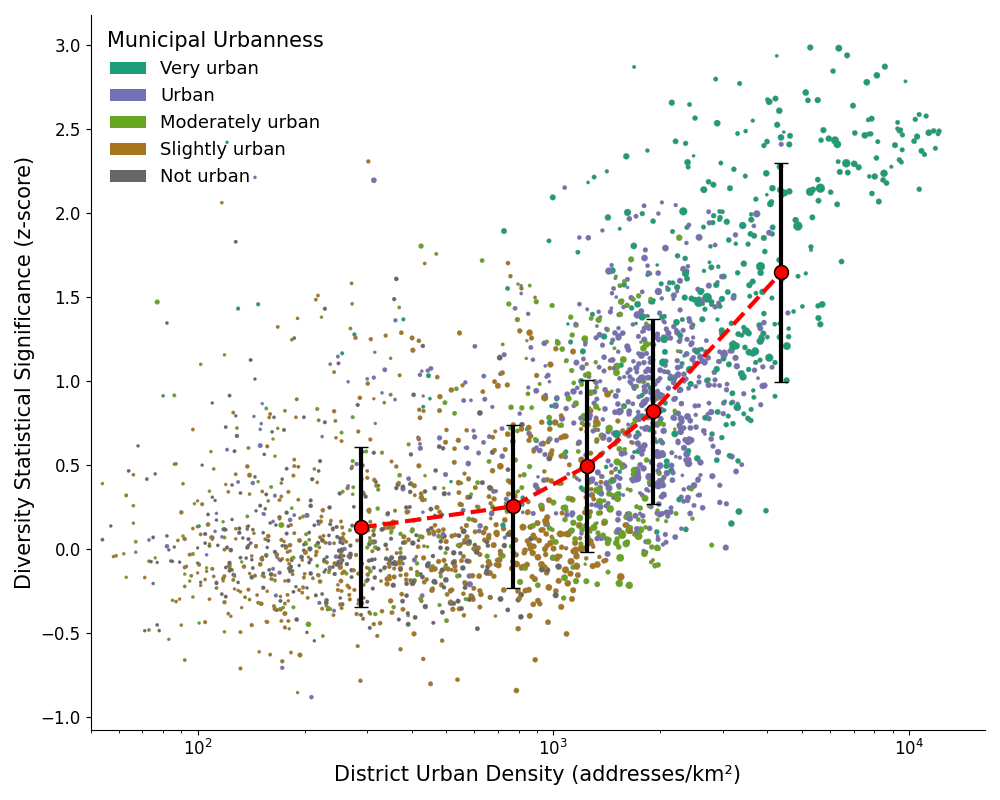}
        \caption{}
        \label{fig:sidi_urban_b}
    \end{subfigure}
    \caption{\textit{Community diversity statistical significance and its link to urbanization. }\textbf{(a)} Geographic heatmap of the $z$-score of community diversity across neighborhoods. Hot spots of high diversity (yellow) are clearly concentrated in major urban areas. \textbf{(b)} Neighborhood diversity statistical significance vs urban density, colored by their parent municipality urbanization level. A strong positive trend indicates that denser, more urban regions are significantly more diverse than what a spatial model alone predicts.}
    \label{fig:sidi_urban}
\end{figure}

To identify the neighborhoods with sufficiently high diversity that is unaccounted for by spatial factors, we compute the statistical significance of the diversity through the $z$-score (see Section~\ref{sec:diversity}). Visualizing the $z$-score on a map (Figure~\ref{fig:sidi_urban_a}) effectively removes the sensitivity to community borders and reveals a clear pattern: significant diversity is concentrated in and around the nation's largest cities. Amsterdam, The Hague, and Rotterdam emerge as prominent hot spots. Notably, Groningen also stands out, exhibiting exceptionally high diversity given its geographically remote location. This may reflect that, being a traditional student city, Groningen attracts young people from across the country. 

This relationship is further quantified in Figure~\ref{fig:sidi_urban_b}, which shows a strong positive correlation between a neighborhood's $z$-score and its population density. This effect is further amplified by the urbanization level of the surrounding municipality. This finding indicates that urban environments act as melting pots, facilitating a level of community mixing that significantly exceeds the baseline interaction predicted by physical proximity. Notably, this metric highlights the unique position of Groningen. Despite its geographic isolation (resulting in low absolute diversity), it exhibits the highest z-score in the country ($z=2.36$, GSI=0.30), identifying it as a highly integrative hub relative to its spatial constraints. It is followed closely by the major cities of the Randstad conurbation, including The Hague ($z=2.35$), Delft ($z=2.15$), and Amsterdam ($z=1.98$). The fact that the method identifies these known urban centers as statistically significant outliers confirms LFMM's capacity to detect complex social connectivity patterns in aggregated data.

\section{Discussion and Conclusion}
\label{sec:discussion}
Our analysis of the population-scale social network of the Netherlands demonstrates that traditional disjoint community detection, while capturing spatial administrative boundaries, fails to represent intermixed aggregated social networks. This confirms that treating aggregated sets as indivisible units risks obscuring the underlying network structure. The proposed Link Fraction Mixed Membership (LFMM) method resolves this by treating an aggregate node as a sum of diversely-connected nodes rather than an atomic unit. By ensuring that membership sums are conserved under aggregation and disaggregation, LFMM provides a consistent link between the macroscale aggregated network and the microscale heterogeneity of the individual-level graph. This consistency allows for a robust analysis that is not strictly bound by the specific resolution or shape of the aggregation partition.

We validated the utility of the method by applying it to the population-scale social network of the Netherlands. While raw mixed membership values largely reflected the network's strong spatial embedding, the integration of a gravity null model allowed us to disentangle diversity arising from geographic proximity from that driven by social preference. This distinction was crucial for identifying that high levels of significant diversity are notably correlated with urbanization. The finding that urban centers function as ``melting pots'' demonstrates that LFMM is capable of detecting subtle, non-spatial structural patterns that are typically dominated by the geographic constraints of the network.

Some caution is necessary, however, when interpreting these mixed membership results. The edge-centric nature of the method offers a distinct perspective. Because the membership vectors are weighted by node strength, they reflect the volume of connectivity within a region rather than the number of residing individuals. While this means that high-degree hubs may disproportionately influence the aggregate profile, it provides a meaningful measure of connectivity between groups. Furthermore, while the conservation properties of LFMM as a mixed membership method hold true for any aggregated network, the method does not distinguish between whether individuals are connected to multiple communities, versus the presence of homogeneously connected members of multiple communities. Consequently, the only way to decouple those two possibilities is to know the level of alignment of the aggregation partition with the community structure at the individual-level 
{as was done with the Dutch network. The high aggregability values suggest that diverse connectivity makes a substantial contribution to the observed community diversity.} 

Future work could focus on benchmarking LFMM against inference-based approaches, such as Mixed Membership Stochastic Block Models (MMSBM), to explore which conceptual definitions of membership are most suitable for different analytical goals. Additionally, the versatility of the method should be tested on a broader range of network types, including directed, weighted, multiplex, and link-partitioned networks. Ultimately, LFMM enables the robust analysis of community composition and dynamics, especially in (spatially) aggregated temporal networks. By uncovering community composition diversity and evolution, it facilitates our understanding of the complex structure and dynamics of communities in large-scale networks.

\subsection*{Funding information}

The authors acknowledge that they received funding in support for this research from the Dutch Research Council (NWO) through the SSHOC-NL project. 

\subsection*{Author contributions}

G.A., E.B., E.M.H., and F.W.T. conceived the study. G.A. performed the data work, developed the code and the methodology, led the analyses, and created the figures. E. B. contributed to the methods and analyses. All authors contributed to writing, reviewed the manuscript, and approved the final version.

\subsubsection*{Data availability statement}

All data needed to evaluate the conclusions in the paper as well as access procedures and further information on the dataset are deposited in the secure storage of the ODISSEI portal\footnote{\href{https://odissei-data.nl/facility/odissei-portal/}{https://odissei-data.nl/facility/odissei-portal/}} in the following repository: \href{https://doi.org/10.34894/8575OP}{https://doi.org/10.34894/8575OP}. Access can be requested after obtaining authorization to use the Statistics Netherlands (CBS) Remote Access (RA) Microdata environment\footnote{\href{https://www.cbs.nl/en-gb/our-services/ customised-services-microdata}{https://www.cbs.nl/en-gb/our-services/customised-services-microdata}}.

\subsection*{Code availability}
The code for the Link Fraction Mixed Membership method and community diversity analysis can be found on \url{https://github.com/g-adel/LFMM-Paper}.

\clearpage
\printbibliography

@article{Ahn2009LinkCR,
  author = {Yong-Yeol Ahn and James P. Bagrow and S. Lehmann},
  doi = {10.1038/nature09182},
  journal = {Nature},
  pages = {761-764},
  title = {Link communities reveal multiscale complexity in networks},
  volume = {466},
  year = {2009}
}

@article{Airoldi2007MixedMS,
  author = {E. Airoldi and D. Blei and S. Fienberg and E. Xing},
  doi = {10.1145/1134271.1134283},
  journal = {Journal of Machine Learning Research},
  pages = {1981-2014},
  title = {Mixed Membership Stochastic Blockmodels},
  volume = {9},
  year = {2007}
}

@article{bokanyi2023anatomy,
  author = {Bok{\'a}nyi, Eszter and Heemskerk, Eelke M and Takes, Frank W},
  doi = {10.1038/s41598-023-36324-9},
  journal = {Scientific Reports},
  number = {1},
  pages = {9209},
  publisher = {Nature Publishing Group UK London},
  title = {The anatomy of a population-scale social network},
  volume = {13},
  year = {2023}
}

@article{butts2009revisiting,
  author = {Butts, Carter T},
  doi = {10.1126/science.1171022},
  journal = {Science},
  number = {5939},
  pages = {414--416},
  publisher = {American Association for the Advancement of Science},
  title = {Revisiting the foundations of network analysis},
  volume = {325},
  year = {2009}
}

@incollection{cazabet2023challenges,
  author = {Cazabet, Remy and Rossetti, Giulio},
  booktitle = {Temporal network theory},
  doi = {10.1007/978-3-030-23495-9_10},
  pages = {185--202},
  publisher = {Springer},
  title = {Challenges in community discovery on temporal networks},
  year = {2023}
}

@misc{cho2014mixed,
  author = {Cho, Yoon-Sik and Ver Steeg, Greg and Galstyan, Aram},
  doi = {10.1609/aaai.v25i1.7952},
  title = {Mixed Membership Blockmodels for Dynamic Networks with Feedback.},
  year = {2014}
}

@article{de2024effect,
  author = {de Jong, Rachel G and van der Loo, Mark PJ and Takes, Frank W},
  doi = {10.1038/s41598-023-50617-z},
  journal = {Scientific Reports},
  number = {1},
  pages = {1156},
  publisher = {Nature Publishing Group UK London},
  title = {The effect of distant connections on node anonymity in complex networks},
  volume = {14},
  year = {2024}
}

@article{Decuyper2018MeasuringTE,
  author = {Yérali Gandica and Adeline Decuyper and C. Cloquet and I. Thomas and J. Delvenne},
  doi = {10.1140/epjds/s13688-020-00223-0},
  journal = {EPJ Data Science},
  title = {Measuring the effect of node aggregation on community detection},
  volume = {9},
  year = {2018}
}

@article{Evans2009LineGL,
  author = {T. Evans and R. Lambiotte},
  booktitle = {Physical review. E, Statistical, nonlinear, and soft matter physics},
  doi = {10.1103/physreve.80.016105},
  journal = {Physical review E, Statistical, nonlinear, and soft matter physics},
  pages = {016105},
  title = {Line graphs, link partitions, and overlapping communities.},
  volume = {80 1 Pt 2},
  year = {2009}
}

@article{jones2021scalable,
  author = {Jones, Timothy and Ward, Owen G and Jiang, Yiran and Paisley, John and Zheng, Tian},
  doi = {10.5705/ss.202022.0411},
  journal = {arXiv preprint 2108.01727},
  title = {Scalable Community Detection in Massive Networks using Aggregated Relational Data},
  year = {2021}
}

@article{Jost2006,
  author = {Jost, Lou},
  doi = {https://doi.org/10.1111/j.2006.0030-1299.14714.x},
  eprint = {https://nsojournals.onlinelibrary.wiley.com/doi/pdf/10.1111/j.2006.0030-1299.14714.x},
  journal = {Oikos},
  number = {2},
  pages = {363-375},
  title = {Entropy and diversity},
  url = {https://nsojournals.onlinelibrary.wiley.com/doi/abs/10.1111/j.2006.0030-1299.14714.x},
  volume = {113},
  year = {2006}
}

@article{kallus2015spatial,
  author = {Kallus, Zsofia and Barankai, Norbert and Szuele, Janos and Vattay, Gabor},
  doi = {10.1371/journal.pone.0126713},
  journal = {PLOS ONE},
  number = {5},
  pages = {e0126713},
  publisher = {Public Library of Science San Francisco, CA USA},
  title = {Spatial fingerprints of community structure in human interaction network for an extensive set of large-scale regions},
  volume = {10},
  year = {2015}
}

@article{kazmina2024socio,
  author = {Kazmina, Yuliia and Heemskerk, Eelke M and Bok\'anyi, Eszter and Takes, Frank W},
  doi = {10.1016/j.socnet.2024.02.005},
  journal = {Social Networks},
  pages = {279--291},
  publisher = {Elsevier},
  title = {Socio-economic segregation in a population-scale social network},
  volume = {78},
  year = {2024}
}

@article{Kim2004GeographicalCG,
  author = {Beom Jun Kim},
  doi = {10.1103/physrevlett.93.168701},
  journal = {Physical Review Letters},
  pages = {168701},
  title = {Geographical coarse graining of complex networks.},
  volume = {93 16},
  year = {2004}
}

@misc{peng2018social,
  title={Social networking big data: Opportunities, solutions, and challenges},
  author={Peng, Sancheng and Yu, Shui and Mueller, Peter},
  journal={Future Generation Computer Systems},
  volume={86},
  pages={1456--1458},
  year={2018},
  publisher={Elsevier}
}

@article{Kuppevelt2020CommunityMC,
  author = {D. V. Kuppevelt and R. Bakhshi and E. Heemskerk and Frank W. Takes},
  doi = {10.1007/s42001-021-00145-5},
  journal = {Journal of Computational Social Science},
  pages = {841 - 860},
  title = {Community membership consistency applied to corporate board interlock networks},
  volume = {5},
  year = {2020}
}

@article{mantzaris2014uncovering,
  author = {Mantzaris, Alexander V},
  doi = {10.1140/epjds/s13688-014-0026-9},
  journal = {EPJ Data Science},
  number = {1},
  pages = {26},
  publisher = {Springer},
  title = {Uncovering nodes that spread information between communities in social networks},
  volume = {3},
  year = {2014}
}

@article{Menyhert2024ConnectivityAC,
  author = {M'arton Menyh\'ert and Eszter Bok\'anyi and R. Corten and E. Heemskerk and Yuliia Kazmina and F. Takes},
  doi = {10.1140/epjds/s13688-025-00522-4},
  journal = {EPJ Data Science},
  pages = {8},
  title = {Connectivity and community structure of online and register-based social networks},
  volume = {14},
  year = {2024}
}

@article{peixoto2015inferring,
  author = {Peixoto, Tiago P},
  doi = {10.1103/physreve.92.042807},
  journal = {Physical Review E},
  number = {4},
  pages = {042807},
  publisher = {APS},
  title = {Inferring the mesoscale structure of layered, edge-valued, and time-varying networks},
  volume = {92},
  year = {2015}
}

@article{peixoto2019bayesian,
  author = {Peixoto, Tiago P},
  doi = {10.1002/9781119483298.ch11},
  journal = {Advances in network clustering and blockmodeling},
  pages = {289--332},
  publisher = {Wiley Online Library},
  title = {Bayesian stochastic blockmodeling},
  year = {2019}
}

@article{prieto2018gravity,
  author = {Prieto Curiel, Rafael and Pappalardo, Luca and Gabrielli, Lorenzo and Bishop, Steven Richard},
  doi = {10.1371/journal.pone.0199892},
  journal = {PlOS ONE},
  number = {7},
  pages = {e0199892},
  publisher = {Public Library of Science San Francisco, CA USA},
  title = {Gravity and scaling laws of city to city migration},
  volume = {13},
  year = {2018}
}

@article{reichardt2006statistical,
  author = {Reichardt, J{\"o}rg and Bornholdt, Stefan},
  doi = {10.1103/physreve.74.016110},
  journal = {Physical Review E},
  number = {1},
  pages = {016110},
  publisher = {APS},
  title = {Statistical mechanics of community detection},
  volume = {74},
  year = {2006}
}

@article{robiglio2025multiscale,
      title={Multiscale patterns of migration flows in Austria: regionalization, administrative barriers, and urban-rural divides}, 
      author={Thomas Robiglio and Martina Contisciani and Márton Karsai and Tiago P. Peixoto},
      year={2025},
      journal = {arXiv preprint 2507.11503},
      url={https://arxiv.org/abs/2507.11503}, 
  doi = {10.48550/arXiv.2507.11503},
  month = {07},
  year = {2025}
}

@article{robinson2009ecological,
  author = {Robinson, William S},
  doi = {10.1093/ije/dyn357},
  journal = {International Journal of Epidemiology},
  number = {2},
  pages = {337--341},
  publisher = {Oxford University Press},
  title = {Ecological correlations and the behavior of individuals},
  volume = {38},
  year = {2009}
}

@article{rosvall2010mapping,
  author = {Rosvall, Martin and Bergstrom, Carl T},
  doi = {10.1371/journal.pone.0008694},
  journal = {PlOS ONE},
  number = {1},
  pages = {e8694},
  publisher = {Public Library of Science San Francisco, USA},
  title = {Mapping change in large networks},
  volume = {5},
  year = {2010}
}

@article{soler2024contacts,
  author = {Soler, Nicol{\'a}s and Heemskerk, Eelke and Kazmina, Yuliia},
  doi = {10.31235/osf.io/axumt},
  publisher = {OSF},
  title = {Contacts in contexts: Measuring intergroup contact opportunities at the population-scale through linked administrative and survey data},
  year = {2024}
}

@article{traag2019louvain,
  author = {Traag, Vincent A and Waltman, Ludo and Van Eck, Nees Jan},
  doi = {10.1038/s41598-019-41695-z},
  journal = {Scientific Reports},
  number = {1},
  pages = {1--12},
  publisher = {Nature Publishing Group},
  title = {From Louvain to Leiden: guaranteeing well-connected communities},
  volume = {9},
  year = {2019}
}

@misc{vanderlaan2022person,
  author = {Jan van~der~Laan},
  title = {A Person Network of the Netherlands},
  type = {\url{https://www.cbs.nl/-/media/_pdf/2022/20/person_network_netherlands.pdf}},
  note = {},
  year = {2022}
}

@article{ward2025bayesian,
  author = {Ward, Owen G and Smith, Anna L and Zheng, Tian},
  doi = {10.5705/ss.202022.0411},
  journal = {arXiv preprint 2506.21353},
  title = {Bayesian Modeling for Aggregated Relational Data: A Unified Perspective},
  year = {2025}
}

@inproceedings{backstrom2006group,
  isbn = {1595933395},
  publisher = {Association for Computing Machinery},
  address = {New York, NY, USA},
  url = {https://doi.org/10.1145/1150402.1150412},
  doi = {10.1145/1150402.1150412},
  title={Group formation in large social networks: membership, growth, and evolution},
  author={Backstrom, Lars and Huttenlocher, Dan and Kleinberg, Jon and Lan, Xiangyang},
  booktitle={Proceedings of the 12th ACM SIGKDD international conference on Knowledge discovery and data mining},
  pages={44--54},
  year={2006}
}

@article{Xing2008ASM,
  author = {E. Xing and Wenjie Fu and Le Song},
  doi = {10.1214/09-aoas311},
  journal = {The Annals of Applied Statistics},
  pages = {535-566},
  title = {A state-space mixed membership blockmodel for dynamic network tomography},
  volume = {4},
  year = {2008}
}

@article{yang2014overlapping,
  author = {Yang, Jaewon and Leskovec, Jure},
  doi = {10.1109/jproc.2014.2364018},
  journal = {Proceedings of the IEEE},
  number = {12},
  pages = {1892--1902},
  publisher = {IEEE},
  title = {Overlapping communities explain core--periphery organization of networks},
  volume = {102},
  year = {2014}
}

@incollection{wong2004modifiable,
  author = {Wong, David WS},
  booktitle = {WorldMinds: geographical perspectives on 100 problems: commemorating the 100th anniversary of the association of American geographers 1904--2004},
  doi = {10.4135/9780857020130.n7},
  pages = {571--575},
  publisher = {Springer},
  title = {The modifiable areal unit problem (MAUP)},
  year = {2004}
}

@article{abbe2018community,
  title={Community detection and stochastic block models: recent developments},
  author={Abbe, Emmanuel},
  journal={Journal of Machine Learning Research},
  volume={18},
  number={177},
  pages={1--86},
  year={2018}
}

\clearpage

\appendix

\setcounter{figure}{0}
\setcounter{table}{0}

\section{Computation and Extension of LFMM}
\label{app:matrix_comp}

\subsection{Single matrix computation of LFMM}

The mixed membership vectors for all aggregated sets can be computed simultaneously using linear algebra. Let $n$ be the number of aggregated
sets and $r$ be the number of communities. We define the community indicator matrix $\mathbf C$ of dimension $n\times r$ as:
\begin{equation}
    C_{ik}=
    \begin{cases}
        1 & \text{if } i\in C'_k,\\
        0 & \text{otherwise}.
    \end{cases}
\end{equation}
Here $i\in C'_k$ means that aggregate node $i$ belongs to community $k$. Let $\mathbf W'$ be the aggregated adjacency matrix of dimension $n\times n$, with weights $w'_{ij}$ as defined in Equation~\ref{eq:lfmm_aggregation}. Its diagonal entries represent the total weight of half-edges within each aggregation set and are included in full. The unnormalized mixed membership matrix $\mathbf M'$, where the element $M'_{ik}$ corresponds to the link weight from set $i$ to community $k$, is then obtained by the matrix multiplication:
\begin{equation}
    \mathbf M'=\mathbf W'\mathbf C.
\end{equation}
To obtain the normalized mixed membership matrix, we define the diagonal strength matrix $\mathbf D$ where $D_{ii}=s'_i=\sum_j w'_{ij}$. The normalized matrix is then:
\begin{equation}
    \mathbf m'=\mathbf D^{-1}\mathbf M'.
\end{equation}
This normalization applies to nodes with positive strength. Nodes with zero strength have zero unnormalized membership but no defined normalized mixed membership and are excluded from the normalized computation.

\subsection{Extension to higher-order diffusion}

The formulation above captures direct connectivity, equivalent to a single step of a random walker. To extend LFMM to account for higher-order connectivity over $t$ discrete steps, we construct the row-stochastic transition matrix $\mathbf P$ using the strength matrix defined above:
\begin{equation}
    \mathbf P=\mathbf D^{-1}\mathbf W'.
\end{equation}
Here, $P_{ij}$ represents the probability that a random walker at node $i$ moves to node $j$ in one step. The normalized mixed membership matrix
after $t$ steps, denoted as $\mathbf m'^{(t)}$, is computed by raising the transition matrix to the power of $t$ and projecting onto the communities:
\begin{equation}
    \mathbf m'^{(t)}=\mathbf P^t\mathbf C.
\end{equation}
In this formulation, the element $m'^{(t)}_{ik}$ represents the probability that a random walker starting at aggregated set $i$ will be located within community $k$ after exactly $t$ steps. The standard normalized LFMM corresponds to the case where $t=1$. This extension describes diffusion on the aggregated network; the consistency under aggregation proven for standard LFMM does not generally hold for $t>1$.
\section{Mixed Membership of Municipalities}
\vspace{-1em}
\begin{table}[H]
\centering
\small
\caption{Mixed membership of top 40 municipalities by population (2021)}
\label{tab:top40_municipalities}
\begin{tabular}{lc*{9}{c}}
\hline
Municipality & Population & \rotatebox{90}{Northern} & \rotatebox{90}{Overijssel} & \rotatebox{90}{N. Holland} & \rotatebox{90}{Utrecht} & \rotatebox{90}{Gelderland} & \rotatebox{90}{S. Holland} & \rotatebox{90}{Brabant} & \rotatebox{90}{Zeeland} & \rotatebox{90}{Limburg} \\
\hline
Amsterdam & 870\,084 & 0.007 & 0.008 & \textbf{0.897} & 0.029 & 0.007 & 0.037 & 0.011 & 0.001 & 0.003 \\
Rotterdam & 651\,188 & 0.004 & 0.005 & 0.024 & 0.018 & 0.005 & \textbf{0.916} & 0.021 & 0.004 & 0.003 \\
's-Gravenhage & 548\,108 & 0.005 & 0.005 & 0.036 & 0.016 & 0.005 & \textbf{0.916} & 0.012 & 0.002 & 0.003 \\
Utrecht & 359\,118 & 0.010 & 0.015 & 0.063 & \textbf{0.818} & 0.015 & 0.045 & 0.025 & 0.004 & 0.006 \\
Eindhoven & 235\,645 & 0.003 & 0.004 & 0.012 & 0.015 & 0.014 & 0.017 & \textbf{0.897} & 0.002 & 0.034 \\
Groningen & 233\,127 & \textbf{0.912} & 0.031 & 0.019 & 0.013 & 0.006 & 0.012 & 0.005 & 0.001 & 0.001 \\
Tilburg & 221\,952 & 0.002 & 0.004 & 0.010 & 0.016 & 0.011 & 0.023 & \textbf{0.914} & 0.004 & 0.017 \\
Almere & 214\,642 & 0.014 & 0.026 & \textbf{0.876} & 0.041 & 0.007 & 0.025 & 0.008 & 0.001 & 0.002 \\
Breda & 184\,045 & 0.004 & 0.005 & 0.017 & 0.021 & 0.010 & 0.054 & \textbf{0.871} & 0.008 & 0.010 \\
Nijmegen & 177\,371 & 0.006 & 0.017 & 0.015 & 0.033 & \textbf{0.846} & 0.014 & 0.049 & 0.002 & 0.018 \\
Apeldoorn & 164\,731 & 0.015 & \textbf{0.803} & 0.025 & 0.050 & 0.069 & 0.020 & 0.011 & 0.002 & 0.003 \\
Haarlem & 162\,517 & 0.008 & 0.007 & \textbf{0.890} & 0.021 & 0.006 & 0.053 & 0.010 & 0.001 & 0.003 \\
Arnhem & 162\,413 & 0.009 & 0.030 & 0.021 & 0.051 & \textbf{0.839} & 0.018 & 0.023 & 0.001 & 0.008 \\
Enschede & 159\,747 & 0.016 & \textbf{0.905} & 0.014 & 0.014 & 0.030 & 0.011 & 0.007 & 0.001 & 0.002 \\
Haarlemmermeer & 157\,762 & 0.007 & 0.007 & \textbf{0.842} & 0.021 & 0.005 & 0.105 & 0.009 & 0.001 & 0.002 \\
Amersfoort & 157\,446 & 0.015 & 0.029 & 0.064 & \textbf{0.829} & 0.016 & 0.030 & 0.014 & 0.002 & 0.003 \\
Zaanstad & 156\,862 & 0.007 & 0.006 & \textbf{0.932} & 0.017 & 0.004 & 0.026 & 0.006 & 0.001 & 0.002 \\
's-Hertogenbosch & 155\,463 & 0.004 & 0.006 & 0.017 & 0.034 & 0.024 & 0.022 & \textbf{0.877} & 0.003 & 0.013 \\
Zwolle & 129\,857 & 0.053 & \textbf{0.840} & 0.030 & 0.033 & 0.017 & 0.017 & 0.007 & 0.001 & 0.002 \\
Zoetermeer & 125\,223 & 0.006 & 0.006 & 0.031 & 0.020 & 0.005 & \textbf{0.915} & 0.013 & 0.002 & 0.002 \\
Leeuwarden & 124\,493 & \textbf{0.914} & 0.022 & 0.027 & 0.013 & 0.005 & 0.012 & 0.005 & 0.001 & 0.001 \\
Leiden & 124\,051 & 0.007 & 0.007 & 0.063 & 0.023 & 0.006 & \textbf{0.876} & 0.013 & 0.003 & 0.003 \\
Maastricht & 120\,212 & 0.003 & 0.003 & 0.013 & 0.010 & 0.008 & 0.012 & 0.032 & 0.001 & \textbf{0.917} \\
Dordrecht & 119\,112 & 0.005 & 0.006 & 0.018 & 0.025 & 0.006 & \textbf{0.874} & 0.057 & 0.006 & 0.004 \\
Ede & 118\,541 & 0.010 & 0.030 & 0.024 & \textbf{0.824} & 0.060 & 0.030 & 0.016 & 0.003 & 0.004 \\
Alphen a.d. Rijn & 112\,616 & 0.007 & 0.008 & 0.056 & 0.038 & 0.006 & \textbf{0.868} & 0.013 & 0.003 & 0.002 \\
Westland & 111\,385 & 0.005 & 0.005 & 0.020 & 0.013 & 0.004 & \textbf{0.937} & 0.013 & 0.003 & 0.002 \\
Alkmaar & 109\,886 & 0.010 & 0.007 & \textbf{0.925} & 0.016 & 0.004 & 0.028 & 0.007 & 0.001 & 0.002 \\
Emmen & 107\,031 & \textbf{0.908} & 0.044 & 0.014 & 0.011 & 0.006 & 0.011 & 0.004 & 0.000 & 0.001 \\
Delft & 103\,578 & 0.006 & 0.007 & 0.039 & 0.021 & 0.006 & \textbf{0.897} & 0.017 & 0.003 & 0.003 \\
Venlo & 101\,968 & 0.002 & 0.004 & 0.008 & 0.009 & 0.026 & 0.009 & 0.053 & 0.001 & \textbf{0.889} \\
Deventer & 101\,223 & 0.018 & \textbf{0.845} & 0.020 & 0.025 & 0.064 & 0.014 & 0.010 & 0.001 & 0.003 \\
Helmond & 92\,629 & 0.002 & 0.004 & 0.009 & 0.011 & 0.015 & 0.013 & \textbf{0.910} & 0.001 & 0.034 \\
Oss & 92\,542 & 0.003 & 0.006 & 0.010 & 0.027 & 0.066 & 0.013 & \textbf{0.863} & 0.001 & 0.010 \\
Sittard-Geleen & 91\,728 & 0.002 & 0.003 & 0.008 & 0.008 & 0.008 & 0.010 & 0.036 & 0.001 & \textbf{0.924} \\
Amstelveen & 90\,824 & 0.006 & 0.007 & \textbf{0.902} & 0.029 & 0.005 & 0.040 & 0.009 & 0.001 & 0.002 \\
\hline
\end{tabular}
\end{table}

\end{document}